\documentclass[final,authoryear,5p,times,twocolumn]{elsarticle} %%, longtitle

\usepackage{graphics}
\usepackage{amssymb}
\usepackage{lineno}

\usepackage{color}
\usepackage{fixltx2e}
\usepackage{aas_macros}
\journal{Astroparticle Physics}

\begin{document}

\begin{frontmatter}

%% Title, authors and addresses

%% use the tnoteref command within \title for footnotes;
%% use the tnotetext command for the associated footnote;
%% use the fnref command within \author or \address for footnotes;
%% use the fntext command for the associated footnote;
%% use the corref command within \author for corresponding author footnotes;
%% use the cortext command for the associated footnote;
%% use the ead command for the email address,
%% and the form \ead[url] for the home page:
%%
%% \title{Title\tnoteref{label1}}
%% \tnotetext[label1]{}
%% \author{Name\corref{cor1}\fnref{label2}}
%% \ead{email address}
%% \ead[url]{home page}
%% \fntext[label2]{}
%% \cortext[cor1]{}
%% \address{Address\fnref{label3}}
%% \fntext[label3]{}

%\title{The major upgrade of the MAGIC telescopes, Part II: The achieved physics performance using the Crab Nebula observations}
\title{\vspace{-2.5cm}The major upgrade of the MAGIC telescopes, Part II: A performance study using observations of the Crab Nebula}

%% use optional labels to link authors explicitly to addresses:
%% \author[label1,label2]{<author name>}
%% \address[label1]{<address>}
%% \address[label2]{<address>}

%%\author{authors...}
%%\address{addresses...}
% authors 02.06.2014  Format elsart
%
\author[a]{\vspace{-0.3cm}J.~Aleksi\'c}
\author[b]{S.~Ansoldi}
\author[c]{L.~A.~Antonelli}
\author[d]{P.~Antoranz}
\author[e]{A.~Babic}
\author[f]{P.~Bangale}
\author[a]{M.~Barcel\'o} 
\author[g]{J.~A.~Barrio}
\author[h,*]{J.~Becerra Gonz\'alez}
\author[i]{W.~Bednarek}
% ...

\address[a]{IFAE, Campus UAB, E-08193 Bellaterra, Spain}
\address[b]{Universit\`a di Udine, and INFN Trieste, I-33100 Udine, Italy}
\address[c]{INAF National Institute for Astrophysics, I-00136 Rome, Italy}
\address[d]{Universit\`a  di Siena, and INFN Pisa, I-53100 Siena, Italy}
\address[e]{Croatian MAGIC Consortium, Rudjer Boskovic Institute, University of Rijeka and University of Split, HR-10000 Zagreb, Croatia}
\address[f]{Max-Planck-Institut f\"ur Physik, D-80805 M\"unchen, Germany}
\address[g]{Universidad Complutense, E-28040 Madrid, Spain}
\address[h]{Inst. de Astrof\'isica de Canarias, E-38200 La Laguna, Tenerife, Spain}
\address[i]{University of \L\'od\'z, PL-90236 Lodz, Poland}
% ...
\address[*]{now at: NASA Goddard Space Flight Center, Greenbelt, MD 20771, USA and Department of Physics and Department of Astronomy, University of Maryland, College Park, MD 20742, USA}
% ...
\cortext[bbb]{Corresponding authors: J. Sitarek (jsitarek@uni.lodz.pl), E. Carmona (emiliano.carmona@ciemat.es), P. Colin (colin@mppmu.mpg.de)}

\begin{abstract}
MAGIC is a system of two Imaging Atmospheric Cherenkov Telescopes located in the Canary island of La Palma, Spain. 
During summer 2011 and 2012 it underwent a series of upgrades, involving the exchange of the MAGIC-I camera and its trigger system, as well as the upgrade of the readout system of both telescopes. 
We use observations of the Crab Nebula taken at low and medium zenith angles to assess the key performance parameters of the MAGIC stereo system. 
For low zenith angle observations, the standard trigger threshold of the MAGIC telescopes  is $\sim50\,$GeV.
The integral sensitivity for point-like sources with Crab Nebula-like spectrum above 220\,GeV is $(0.66\pm0.03)\%$ of Crab Nebula flux in 50\,h of observations.
The angular resolution, defined as the $\sigma$ of a 2-dimensional Gaussian distribution, at those energies is $\lesssim0.07^\circ$, while the energy resolution is 16\%.
We also re-evaluate the effect of the systematic uncertainty on the data taken  with the MAGIC telescopes after the upgrade. 
We estimate that the systematic uncertainties can be divided in the following components: $< 15\%$ in energy scale, 11-18\% in flux normalization and $\pm0.15$ for the energy spectrum power-law slope.
\end{abstract}

\begin{keyword}
Gamma-ray astronomy \sep Cherenkov telescopes \sep Crab Nebula
%% keywords here, in the form: keyword \sep keyword

%% MSC codes here, in the form: \MSC code \sep code
%% or \MSC[2008] code \sep code (2000 is the default)

\end{keyword}

\end{frontmatter}

%\linenumbers

%% main text
%% %-----------------------------------------------------------------------------
\section{Introduction}\label{intro}
MAGIC (Major Atmospheric Gamma Imaging Cherenkov telescopes) consists of two 17~m diameter Imaging Atmospheric Cherenkov Telescopes (IACT). 
The telescopes are located at a height of 2200 m a.s.l. on the Roque de los Muchachos Observatory on the Canary Island of La Palma, Spain ($28^\circ$N, $18^\circ$W).
They are used for observations of particle showers produced in the atmosphere by very high energy (VHE, $\gtrsim30\,$GeV) $\gamma$-rays. 
Both telescopes are normally operated together in the so-called stereoscopic mode, in which only events seen simultaneously in both telescopes are triggered and analyzed \citep{magic_stereo}. 

Between summer 2011 and 2012 the telescopes went through a major upgrade, carried out in two stages. 
In summer 2011 the readout systems of both telescopes were upgraded.
The multiplexed FADCs used before in MAGIC-I \citep{magic_mux} as well as the Domino Ring Sampler version 2 used in MAGIC-II (DRS2, \citealp{magic_daq}) have been replaced by  Domino Ring Sampler version 4 chips (DRS4, \citealp{ritt_drs4,magic_drs4}).
Besides lower noise, the switch to DRS4 based readout allowed to eliminate the $\sim 10\%$ dead time present in the previous system due to the DRS2 chip. 
In summer 2012 the second stage of the upgrade followed with an exchange of the camera of the MAGIC-I telescope to a uniformly pixelized one \citep{magic_upgrade}. 
The new MAGIC-I camera is equipped with 1039 photomultipliers (PMTs), identical to the MAGIC-II telescope.
Each of the camera pixels covers a field of view of $0.1^\circ$, resulting in a total field of view of $\sim3.5^\circ$.
The upgrade of the camera allowed to increase the area of the trigger region in MAGIC-I by a factor of 1.7 to the value of ${4.8^\circ}^2$. 
In the first part of this article \citep{mup_part1} we described in detail the hardware improvements and the commissioning of the system.
In this second part we focus on the performance of the upgraded system based on Crab Nebula observations. 

The Crab Nebula is a nearby ($\sim1.9$~kpc away, \citealp{tr73}) pulsar wind nebula, and the first source detected in VHE $\gamma$ rays \citep{whipple_crab}.
A few years ago, the satellite $\gamma-$ray telescopes, AGILE \& Fermi-LAT observed flares from the Crab Nebula at GeV energies \citep{ta11, ab10}.
However so far no confirmed variability in the VHE range was found. 
Therefore, since the Crab Nebula is still considered the brightest steady VHE $\gamma$-ray source, it is commonly referred to as the ``standard candle'' of VHE $\gamma$-ray astronomy, and it is frequently used to evaluate the performance of VHE instruments. 
In this paper we use Crab Nebula data to quantify the improvement in performance of the MAGIC telescopes after the aforementioned upgrade.
In Section~\ref{sec:data} we describe the different samples of Crab Nebula data used in the analysis.
In Section~\ref{sec:analysis} we explain the techniques and methods used for the processing of the MAGIC stereo data.
In Section~\ref{sec:results} we evaluate the performance parameters of the MAGIC telescopes after the upgrade. 
In Section~\ref{sec:systematics} we discuss the influence of the upgrade on the systematic uncertainties of the measurements and quantify them. 
The final remarks and summary are gathered in Section~\ref{sec:concl}.

%% %-----------------------------------------------------------------------------
\section{Data sample}\label{sec:data}

In order to evaluate the performance of the MAGIC telescopes, we use several samples of Crab Nebula data taken in different conditions between October 2013 and January 2014.
Notice that, as MAGIC is located in the Northern Hemisphere, the Crab Nebula is observable only during the winter season. 
The data were taken in the standard L1-L3 trigger condition (see \citealp{mup_part1}).
The data selection was mostly based on zenith angle dependent rate of background events surviving the stereo reconstruction.
Other measurements: LIDAR information, observation logbook, daily check of weather and hardware status \citep{go13} are also used as auxiliary information.
All data have been taken in the so-called wobble mode \citep{fo94}, i.e. with the source position offset by a fixed angle, $\xi$, from the camera center in a given direction. 
This method allows to estimate the background from other positions in the sky at the same offset $\xi$. 
Most of the results are obtained using the data taken at low zenith angles ($<30^\circ$) and with the standard wobble offset of $0.4^\circ$.
To evaluate the performance at higher zenith angles we use a medium zenith angle sample ($30-45^\circ$). 
In addition, several low zenith angle samples taken at different offsets are used to study the sensitivity for off-axis observations.
All the data samples are summarized in Table~\ref{tab:data}.

\begin{table}[h]
\centering
\small
\begin{tabular}{r|l|r}
zenith angle [$^\circ$] & offset $\xi [^\circ]$ & time[h] \\\hline\hline
0-30 & 0.40 & 11.1 \\ \hline
30-45& 0.40 & 4.0 \\ \hline
0-30 & 0.20 & 2.3 \\ \hline
0-30 & 0.35 & 0.9 \\ \hline
0-30 & 0.70 & 1.9 \\ \hline
0-30 & 1.00 & 4.2 \\ \hline
0-30 & 1.40 & 4.6 \\ \hline
0-30 & 1.80 & 4.1 \\ \hline
\end{tabular}
\caption{
Zenith angle range, wobble offset angles $\xi$, and effective observation time of the Crab Nebula samples used in this study.
}\label{tab:data}
\end{table}

In addition, to analyze the data and to evaluate some of the performance parameters, such as the energy threshold or the energy resolution we used Monte Carlo (MC) simulations. 
The MC simulations were produced with the standard MAGIC simulation package \citep{ma05}, with the gamma-ray showers generated using the Corsika code \citep{he98}.

%% %-----------------------------------------------------------------------------
\section{Data analysis}\label{sec:analysis}

The data have been analyzed using the standard MAGIC tools: MARS (MAGIC Analysis and Reconstruction Software, \citealp{magic_mars}).
Here we briefly describe all the stages of the standard analysis chain of the MAGIC data.

\subsection{Calibration}
Each event recorded by the MAGIC telescopes consists of the waveform observed in each of the pixels. 
The waveforms span $30\,$ns and are sampled at a frequency of 2Gsamples/s.
The telescopes are triggered with a typical stereo rate of $250-300\,$Hz.
In the first stage of the analysis the pixel signals are reduced to two numbers: charge and arrival time.
The signal is extracted with a simple and robust ``sliding window'' algorithm \citep{magic_fadc}, by finding the maximal integral of 6 consecutive time slices (corresponding to $3\,$ns) within the total readout window. 
The conversion from integrated readout counts to photoelectrons (phe) is done using the F-Factor (excess noise factor) method \citep[see e.g.][]{ml97}.
On average, one phe generates a signal of the order of $\sim$100 integrated readout counts. 
For typical observation conditions the electronic noise and the light of the night sky with such an extractor result in a noise RMS level of $\sim 1\,$phe and a bias (for very small signals) of $\sim 2\,$phe. 
The DRS4 readout requires some special calibration procedures, such as the correction of the time inhomogeneity of the domino ring \citep[see][for details]{magic_drs4} which are applied at this stage.
A small fraction of channels (typically $<1\%$) might be malfunctioning, or be affected by bright stars in their field of view. 
The charge and time information of these pixels are interpolated from their neighbouring ones if at least 3 neighbouring pixels are valid. 

\subsection{Image cleaning and parametrization}
After the upgrade, the camera of each MAGIC telescope has 1039 pixels, however the Cherenkov light of a typical air shower event illuminates only of the order of 10 pixels. 
Most of the pixel signals are solely induced by the night sky background (NSB) and the electronic noise.
In order to remove pixels containing only noise to obtain the image of the shower we perform the so-called sum image cleaning \citep[cf.][]{ri09, magic_pulsar}.
In the first step, we determine the so-called core pixels. 
For this, we search for compact groups of 2, 3 or 4 neighbouring pixels (2NN, 3NN, 4NN), with a summed charge above a given threshold.
In order to protect against a large signal in a single pixel (e.g. due to an afterpulse) dominating a 3NN or 4NN group the signals are clipped before summation.
The signals in those pixels should arrive within a given time window.
These time windows were optimized using time resolution of the signal extraction \citep[see][]{magic_drs4}.
For pulses just above the charge thresholds, the coincidence probability of signals from showers falling within the time window is $\approx 80-90\%$ for a single 2NN, 3NN or 4NN compact group.
The charge thresholds on the sum of 2NN, 3NN and 4NN groups are $2\times 10.8\,$phe, $3\times 7.8\,$phe, $4\times 6\,$phe and the corresponding time windows: 0.5, 0.7 and $1.1\,$ns.
The values of the charge thresholds were optimized to assure that the probability of an event composed of only NSB and electronic noise to survive the cleaning is $\lesssim6\%$.
This translates directly to the maximum fraction of images affected by spurious islands due to noise fluctuations.
In the second step, boundary pixels are looked for to reconstruct the rest of the shower image.
We loop over all the pixels which have a neighbouring core pixel, and include them into the image if the charge of such boundary pixel lies above $3.5\,$phe and its signal arrives within $1.5\,$ns with respect to this core pixel.
After the upgrade all the charge and time threshold values are the same in both telescopes.

\subsection{Stereo reconstruction}
Only events which survive the image cleaning in both telescopes, amounting to about $80\%$ for standard trigger conditions, are retained in the analysis. 
Afterwards, the events from both telescopes are paired, and a basic stereo reconstruction is performed.
The tentative reconstructed direction of the event is computed from the crossing point from the main axes of the Hillas ellipses \citep{hi85}. 
This first stereo reconstruction provides additional event-wise parameters such as impact (defined as the distance of the shower axis to the telescope position, impact$_1$ and impact$_2$ are computed with respect to the MAGIC-I and MAGIC-II telescopes respectively) and the height of the shower maximum.  

\subsection{$\gamma$/hadron separation}
Most of the events registered by the MAGIC telescopes are cosmic-rays showers, which are mainly of hadronic origin.
Even for a bright source such as the Crab Nebula, the fraction of $\gamma$-ray events in the raw data is only of the order of $10^{-3}$.
The rejection of the hadronic background is done on the basis of image shape information and reconstructed direction.
The $\gamma$/hadron separation is performed with the help of the so-called Random Forest (RF) algorithm \citep{magic_rf}.
It allows to combine, in a straight-forward way, the image shape parameters, the timing of the shower, and the stereo parameters into a single classification parameter, $Hadronness$. 

\begin{figure*}[t]
\centering 
\includegraphics[width=0.33\textwidth]{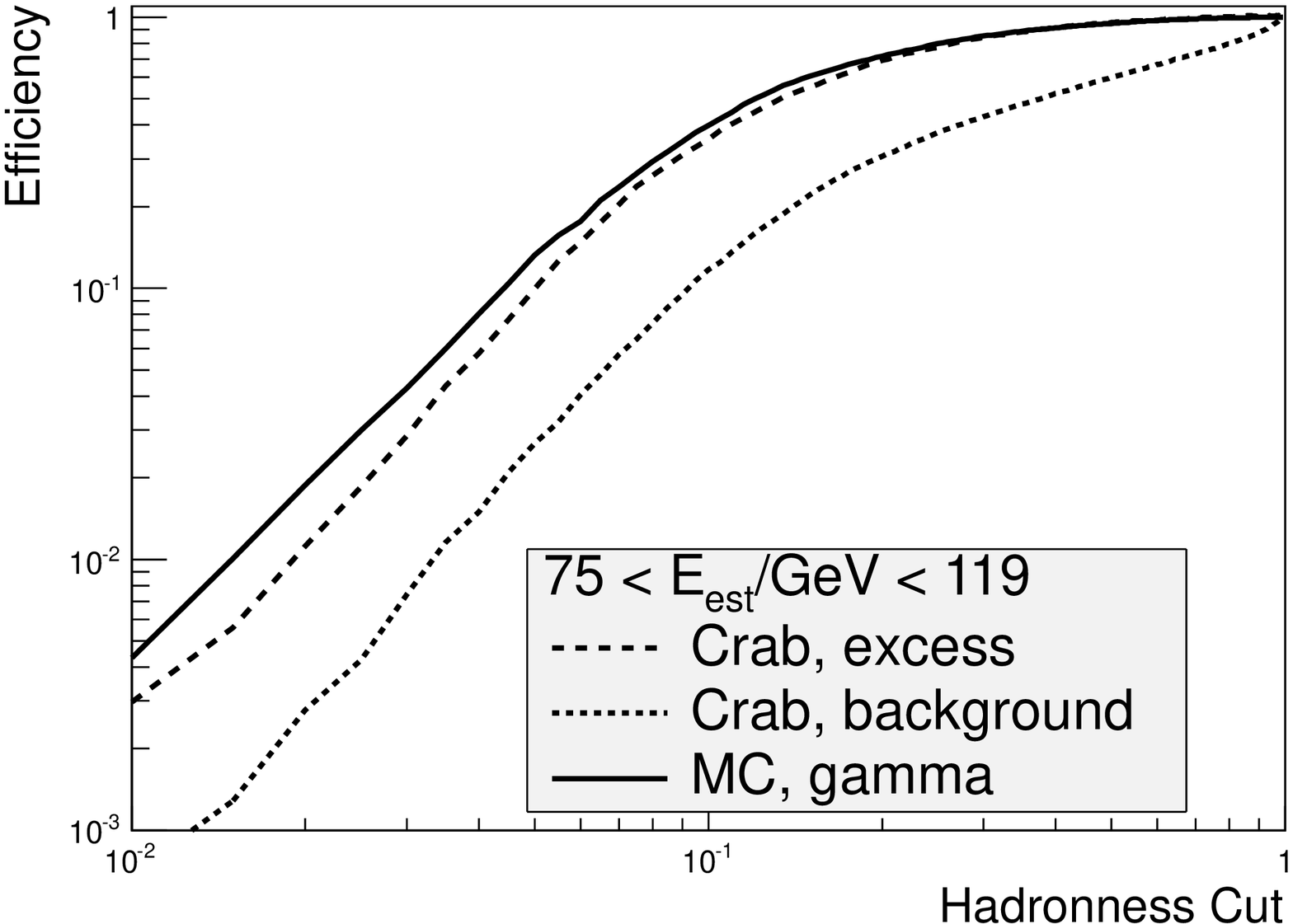}
\includegraphics[width=0.33\textwidth]{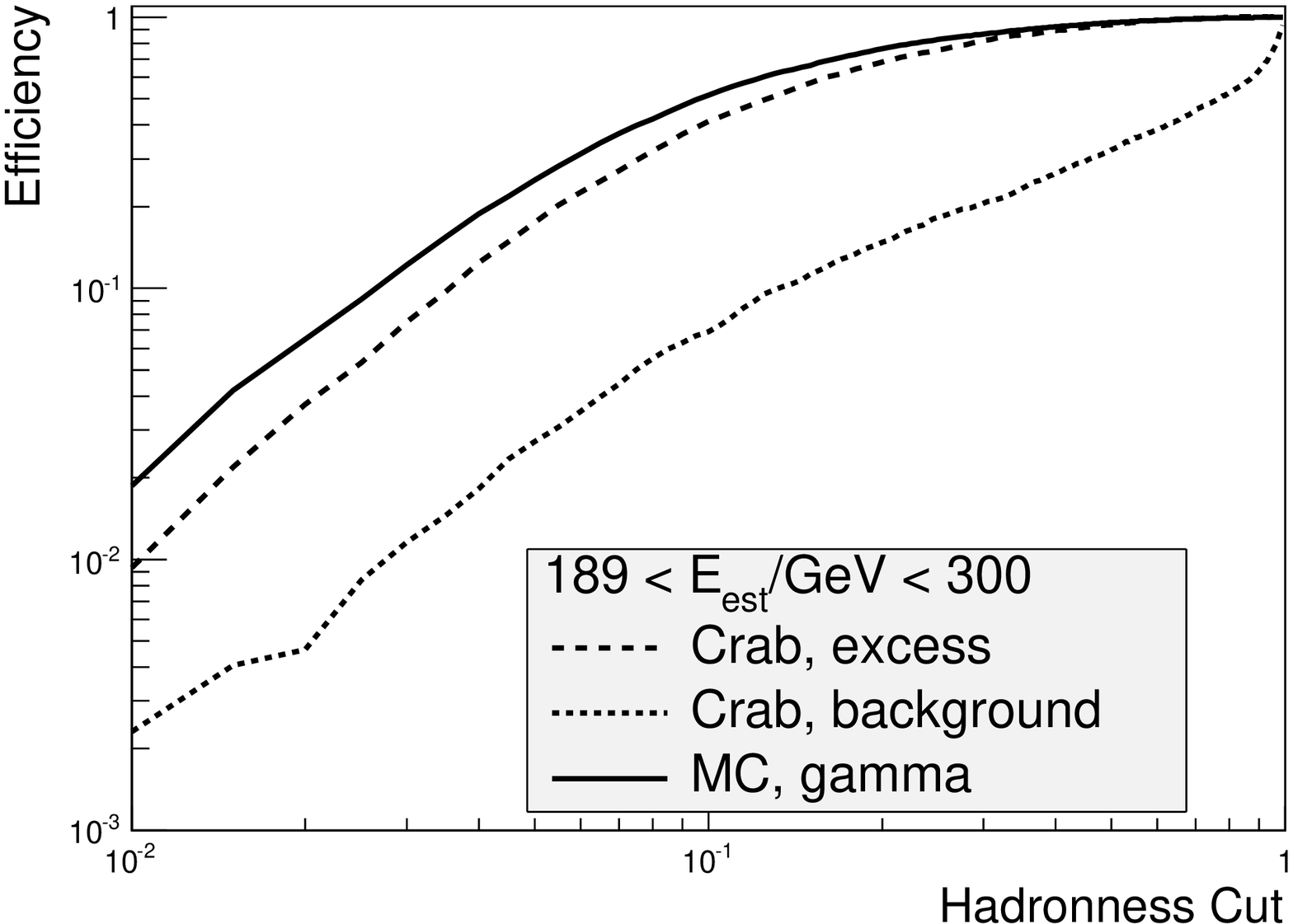}
\includegraphics[width=0.33\textwidth]{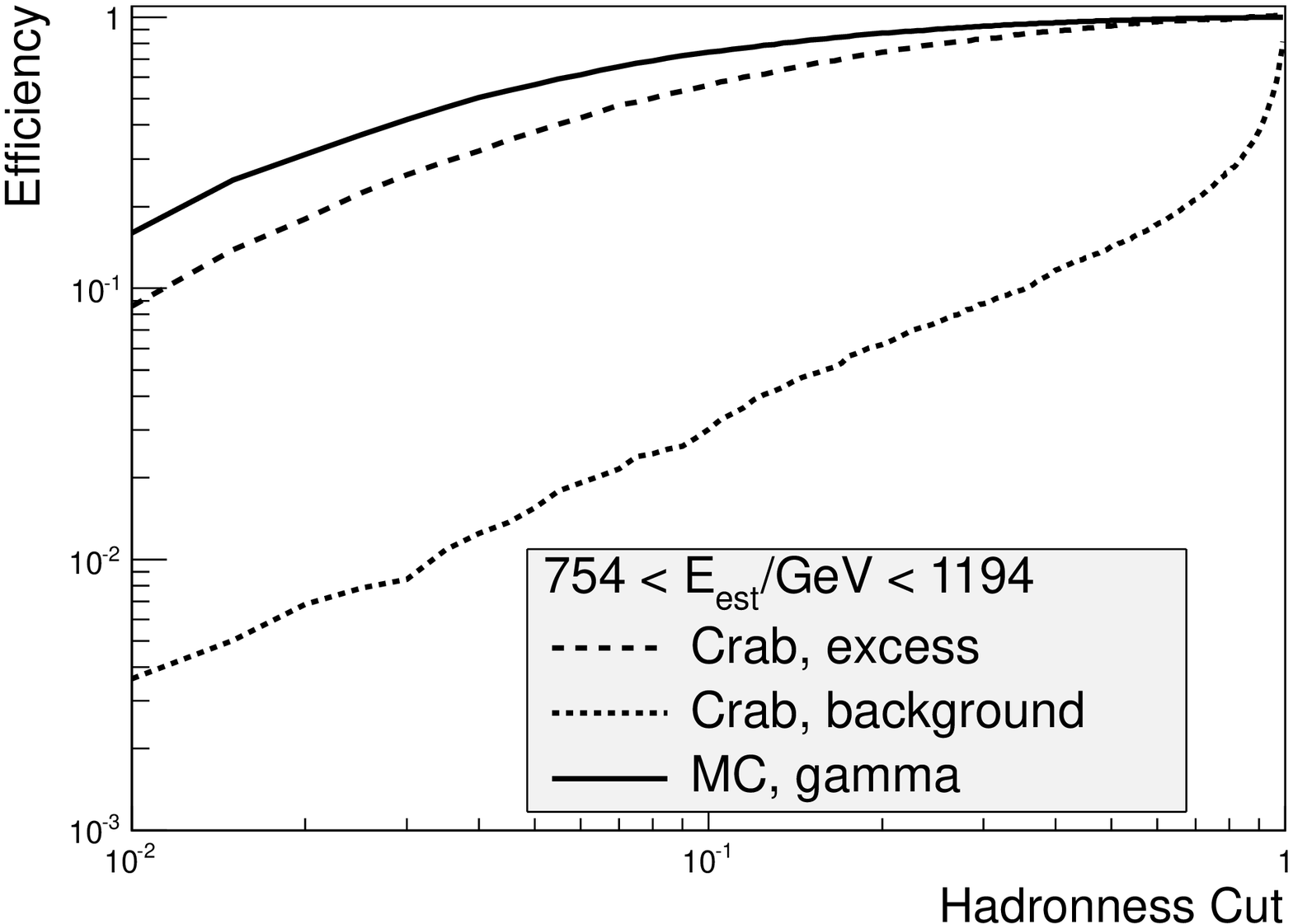}
\caption{
Fraction of events (efficiency) of a given kind: excess $\gamma$-rays (dashed), MC simulated $\gamma$-ray (solid) and background events (dotted) surviving a $Hadronness$ cut.
Different panels correspond to different bins in estimated energy: 75-119\,GeV (left), 189-300\,GeV (center) and 754-1194\,GeV (right).
Additional cuts of $\theta^2<0.03$ and \emph{Size}$>50$\,phe have been applied beforehand. 
}\label{fig:hadr}
\end{figure*}

The survival probability for $\gamma$ rays and background events after a cut in  $Hadronness$ is shown in Fig.~\ref{fig:hadr} for three different energy bins. 

A background rejection better than 90\% can be achieved, with only a small loss of $\gamma$ events.
The $\gamma$/hadron separation performs better at higher energies due to the larger and better defined images.

Strong cuts (below $0.1-0.2$) in $Hadronness$ result in a slight mismatch of the $\gamma$-ray efficiency obtained from the MC simulations with respect to the one from the data itself, which might lead to underestimations in the flux and spectra of the sources.
Consequently, in the process of determining light curves and source spectra, relatively loose cuts in $Hadronness$ are used in order to have a gamma MC efficiency above 90\%, which ensures that the MC-data mismatch in the effective area is below 12\% at the highest energies and below 6\% at the lowest energies.
Note that in addition to this $Hadronness$ cut, the direction reconstruction method (described in subsection~\ref{sec:disp}) also provides additional background suppression. 
In those cases where the accurate determination of the effective area is not relevant (e.g. the analysis with the aim of detecting a source), stronger cuts, which give a better sensitivity, can be used.

\subsection{Arrival direction reconstruction}\label{sec:disp}
\begin{figure}[t]
\centering 
\includegraphics[width=0.49\textwidth]{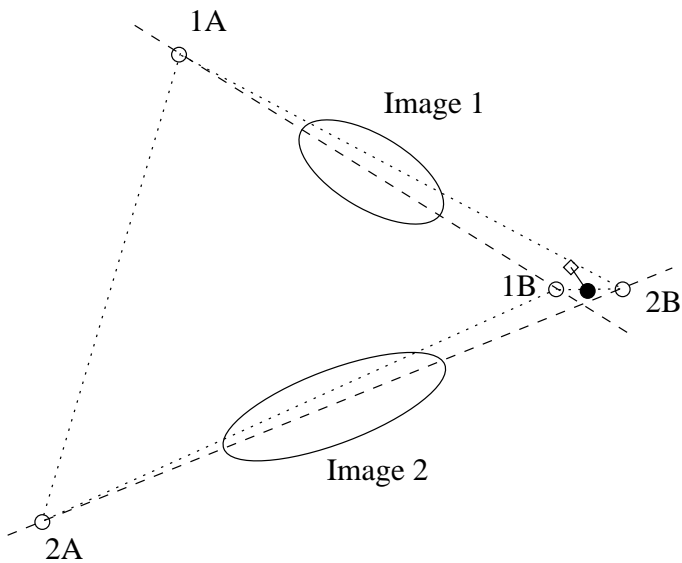}
\caption{
Principle of the Stereo DISP RF method.
The main axes of the images are plotted with dashed lines.
The two DISP RF reconstructed positions per telescope (1A, 1B, 2A, 2B) are shown with empty circles.
The 4 angular distances (1A-2A, 1A-2B, 1B-2A, 1B-2B) are shown with dotted lines. 
The final reconstructed position (the filled circle) is a weighted average of the two closest '1' and '2' points. 
The true source position is marked with a diamond.
}\label{fig:disprf}
\end{figure}

The classical method for arrival direction reconstruction uses the crossing point of the main axes of the Hillas ellipses in the individual cameras \citep{ah97, ho99}.
In the standard MAGIC analysis the event-wise direction reconstruction of the incoming $\gamma$ ray is performed with a DISP RF method.
This method takes into account image shape and timing information, in particular the time gradient measured along the main axis of the image \citep{magic_halo}.
For each telescope we compute an estimated distance, DISP, between the image centroid and the source position. 
As the source position is assumed to be on the line containing the main axis of the Hillas ellipse this results in two possible solutions on either side of the image centroid (see Fig.~\ref{fig:disprf}).
In general, the ambiguity (the so-called head-tail discrimination) can be solved from the asymmetry along the main axis of the image \citep{magic_disp} or from the crossing point of the images. 
However at the lowest energies, where the images consist of few pixels the head-tail discrimination may fail at least in one telescope.
The head-tail discrimination based on the crossing point may also fail,  in the case of close to parallel events. 
Therefore we use a more robust method. 
We compute the 4 distances between the 2 reconstructed positions from each of the telescopes (see dotted lines in Fig.~\ref{fig:disprf}).
We then select the pair of reconstructed positions which give the smallest distance (in the above example 1B-2B).
As the estimation of the DISP parameter is trained with simulated $\gamma$ rays it often gives non-consistent results for hadronic background events.
This provides an extra $\gamma$/hadron separation criterion. 
If none of the four pairs give a similar arrival direction in both telescopes (namely the lowest distance is larger than $0.22^\circ$) the event is discarded.
With this method the fraction of failed head-tail discrimination is between 10\% (at low energies) and $<1\%$ (at high energies).
After determining the correct pair of points, the reconstructed source position is computed as the average of the positions from both telescopes  weighted with the number of pixels in each image. 
The angular distance from this point to the assumed source position is called $\theta$.

\begin{figure}[t]
\centering 
\includegraphics[width=0.49\textwidth]{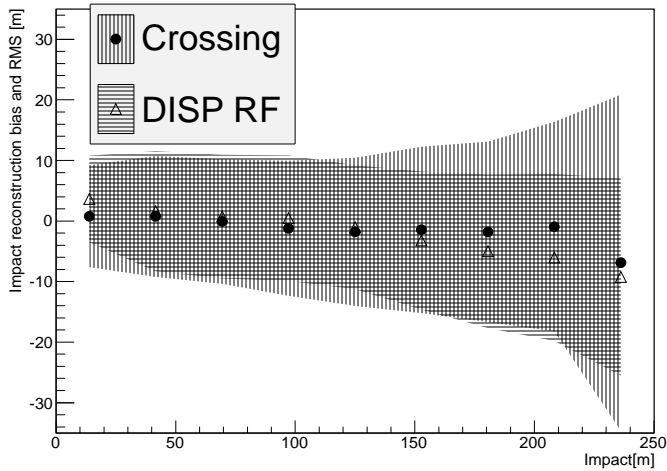}
\caption{
The difference between reconstructed and true impact parameter for events with energy 300-500$\,$GeV.
The markers show the bias (mean of Gaussian) in the reconstruction, and the shaded region the resolution (RMS of Gaussian).
Filled circles and vertical lines show the classical reconstruction based on the crossing point of the images.
Empty triangles and horizontal lines show the reconstruction based on DISP RF method. 
Only events with size $>50$\,phe and $\theta^2<0.02$ are used.
}\label{fig:impact}
\end{figure}

The DISP RF method explained above improves not only the reconstruction of the arrival direction but also the estimation of other shower parameters.
As an example Fig.~\ref{fig:impact} shows the difference between the reconstructed and the true impact parameter for MC $\gamma-$rays with energies of few hundred of GeV. 
The impact parameter can be reconstructed with a precision of about $10\,$m.
The reconstruction with the DISP method is clearly superior for larger values of impact, where a good reconstruction of this parameter for events seen outside of the light pool of the shower is important for improving the energy resolution.

\subsection{Energy estimation}
The event-wise energy estimation is a two-step process.
Using simulated $\gamma$ rays, we build look-up tables relating the event energy to the impact and Cherenkov photon density measured by each telescope (see \cite{magic_stereo} for a detailed explanation).
The look-up tables are based on a simple air-shower Cherenkov emission model that does not reproduce perfectly all the dependencies.
To correct for a zenith angle dependent bias in the energy reconstruction, mainly due to atmospheric absorption, an empirical formula is applied. 
A second correction is applied to account for a small azimuth dependence (due to the geomagnetic field effect).
Finally, a third correction improves the energy reconstruction for large images only partially contained in the camera. 
The final estimated energy, $E_{est}$, is computed as the average of the energies reconstructed individually for each telescope, weighted by the inverse of their uncertainties.

%% %-----------------------------------------------------------------------------
\section{MAGIC performance}\label{sec:results}
In this section we evaluate the main performance parameters of the MAGIC telescopes after the exchange of the readout systems and MAGIC-I camera and compare them with the values from before the upgrade. 

\subsection{Energy threshold}
\begin{figure}[t]
\centering 
\includegraphics[width=0.49\textwidth]{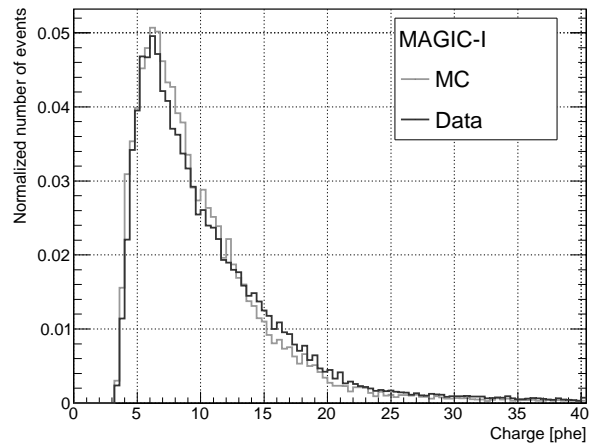}
\includegraphics[width=0.49\textwidth]{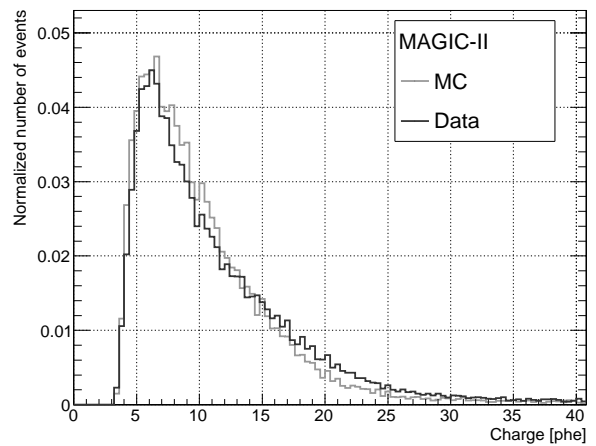}
\caption{
Distribution of the smallest charge in the largest triple of pixels for data (black) and MC protons simulated with a trigger conditions of 3NN above a threshold of $3.9\,$phe (MAGIC-I) or $4.1\,$phe (MAGIC-II) (gray).
Top panel shows MAGIC-I, bottom MAGIC-II. 
}\label{fig:dts}
\end{figure}

The energy threshold of an IACT cannot be obtained in a straight-forward way from the data itself. 
One needs to rely on the Monte Carlo simulations and to make sure that they describe the data appropriately. 
The energy threshold depends on the trigger settings for a given observations.
In particular it depends on an amplitude threshold of individual pixels in the 3NN multiplicity trigger. 
In order to fine tune the trigger parameters in the MC simulations we follow the method described in \cite{magic_stereo}.
For each event that survived the trigger we search for the 3NN combination with the highest signal.
For small showers, just above the threshold, this is the most probable triple which gave a trigger. 
The pixel in this triple with the lowest signal provides a handle for the trigger threshold.
Using many events we build a distribution of such lowest signals in the highest triple.
We then compare the position of the peak with a similar distribution produced from MC protons (see Fig.\ref{fig:dts}). 
We obtain that the individual pixel thresholds are $\approx 3.9$\,phe for MAGIC-I and $\approx 4.1$\,phe for MAGIC-II.
Those values are within 10-15\% consistent with the threshold values obtained directly from the data with an independent method of rate-scans in \cite{mup_part1}.

\begin{figure}[t]
\centering 
\includegraphics[width=0.49\textwidth]{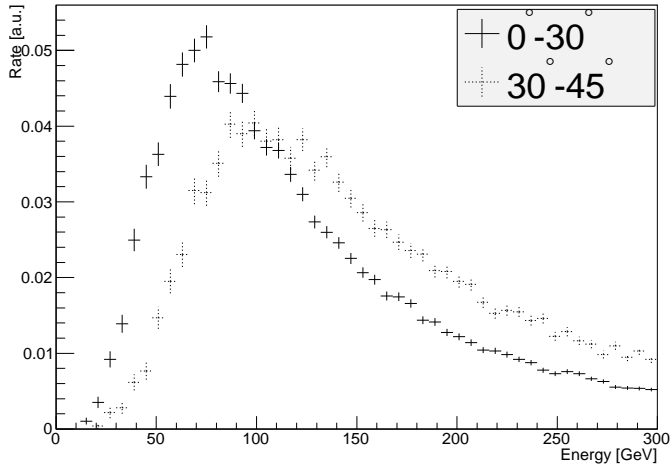}
\caption{
Rate of MC $\gamma-$ray events (in arbitrary units) surviving the image cleaning  with at least 50 phe for a source with a spectral index of $-2.6$.
Solid line: zenith distance below $30^\circ$, dotted line: zenith distance between $30^\circ$ and $45^\circ$.
}\label{fig:threshold_rate}
\end{figure}

In order to study the energy threshold of the MAGIC telescopes we construct a differential rate plot using MC simulations. 
A common definition of an energy threshold is a peak energy of such a plot for a hypothetical source with a spectral index of $-2.6$.
In Fig.~\ref{fig:threshold_rate} we show the differential rate plot in two zenith angle ranges for events that survived image reconstruction in both telescopes.
For low zenith angle, i.e. $<30^\circ$, the reconstruction threshold energy is $\sim 70\,$GeV.
Note however that the peak is broad and extends far to lower energies.
Therefore it is also possible to evaluate the performance of the telescopes and obtain scientific results below such defined threshold.

\begin{figure}[t]
\centering 
\includegraphics[width=0.49\textwidth]{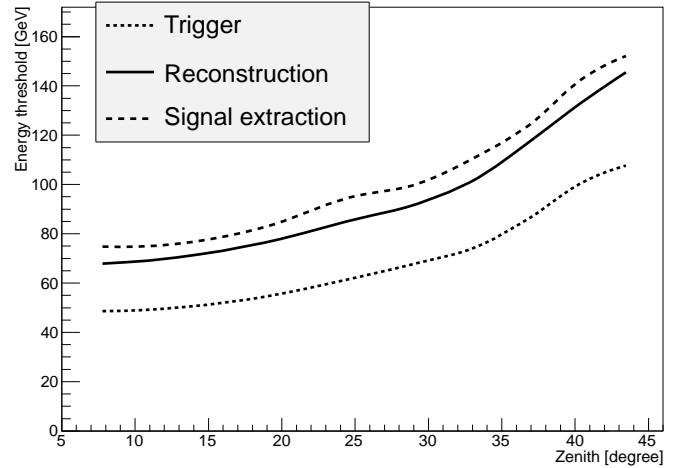}
\caption{
Threshold of the MAGIC telescopes as a function of the zenith angle of the observations.
The energy threshold is defined as the peak energy in the differential rate plot for a source with -2.6 spectral index.
Dotted curve: threshold at the trigger level.
Solid line: only events with images that survived image cleaning in each telescope with at least 50\,phe. 
Dashed line: with additional cuts of $Hadronness<0.5$ and $\theta^2<{0.03^\circ}^2$ applied.
}\label{fig:threshold_vs_zd}
\end{figure}

In Fig.~\ref{fig:threshold_vs_zd} we show the energy threshold of the MAGIC telescopes as the function of zenith distance of observations.
The threshold value is determined by fitting a Gaussian distribution in a narrow range around the peak.
The threshold is quite stable for low zenith angle observations.
It increases rapidly for higher zenith angles, due to larger absorption of the Cherenkov light in the atmosphere and dilution of the photons reaching the ground over a larger light pool.

The threshold can be evaluated at different stages of the analysis.
The trigger threshold computed from all the events that triggered both telescopes is naturally the lowest one, being $\sim 50$\,GeV at low zenith angles.
The shower reconstruction procedure involving image cleaning  and a typical data quality cut of having at least 50 phe in each telescope raises the threshold to $\sim 70$\,GeV.
The events with size lower than this are very small, subjected to high Poissonian fluctuations and therefore harder to reconstruct.
Also the separation of $\gamma$ candidates from the much more abundant hadronic background becomes harder at lower image sizes. 
Signal extraction cuts (the so-called $Hadronness$ cut, and a cut in the angular distance to the nominal source position, $\theta$) increase the threshold further to about 75\,GeV at low zenith angles. 
The value of the energy threshold doubles at zenith angle of $43^\circ$.
In the investigated zenith angle range the value of the threshold after all cuts can be approximated by an empirical formula: $74\times \cos(Zenith\_Angle)^{-2.3}\,\mathrm{GeV}$.

\subsection{Effective collection area}
\begin{figure}[t]
\centering 
\includegraphics[width=0.49\textwidth]{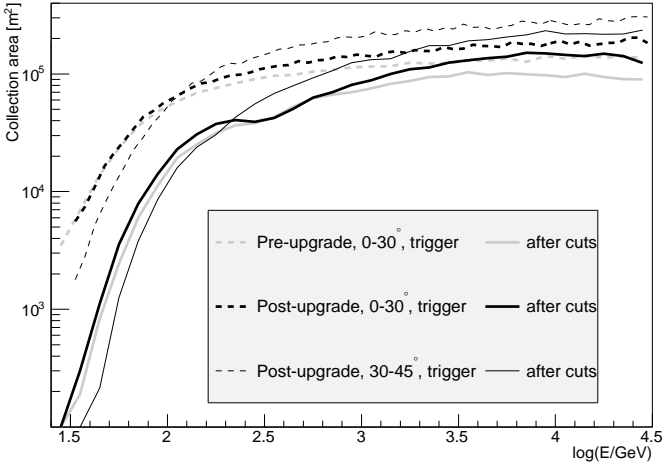}
\caption{
Collection area of the MAGIC telescopes after the upgrade at the trigger level (dashed lines) and after all cuts (solid lines).
Thick lines show the collection area for low zenith angle observations, while thin lines correspond to medium zenith angle. 
For comparison, the corresponding pre-upgrade collection areas are shown with gray lines.
}\label{fig:aeff}
\end{figure}

For large arrays of IACTs the collection area well above the energy threshold for low zenith angle observations is approximately equal to the physical size of the array \citep{cta_mc}. 
On the other hand for a single telescope or small arrays such as the MAGIC telescopes, the collection area is mainly determined by the size of the Cherenkov light pool (radius of $\sim 120\,$m). 
We compute the collection area as the function of the energy $E$ following the standard definition of  $A_{\rm eff}(E) = N(E)/N_{0}(E) \times \pi r_{\max}^2$. 
$N_{0}(E)$ is the number of simulated events, $r_{\max}$ is the maximum simulated shower impact and $N(E)$ is the number of events surviving either the trigger condition or a given set of cuts.
When computing the collection area in broad bins of energy we use weights to reproduce a given spectral shape.
The collection area of the MAGIC telescopes at the trigger level is about $10^5\mathrm{\,m^{2}}$ for 300\,GeV gamma rays (see Fig.~\ref{fig:aeff}). 
In the TeV range it grows slowly with energy, as some of the large showers can be still caught at large values of impact where the density of the Cherenkov photons on the ground falls rapidly. 
Around and below the energy threshold the collection area falls rapidly, as only events with a significant upward fluctuation of the light yield can trigger the telescope. 
At the energy of a few TeV, the trigger collection area after the upgrade is larger by $\sim 30\%$, mostly due to the larger trigger area in the M1 camera.   
The collection area for observations at higher zenith angles is naturally smaller below $\sim 100\,$GeV due to a higher threshold of the observations. 
However, at TeV energies it is larger by $\sim 40\%$ due to an increase of the size of the light pool.

In Fig.~\ref{fig:aeff} we also show the collection area after image cleaning, quality and signal extraction cuts optimized for best differential sensitivity (see Section~\ref{sec:sens}). 
The feature of a dip in the collection area after cuts around $300\,$GeV is caused by a stronger $Hadronness$ cut. 
At those energies the $\gamma$/hadron separation is changing from based on height of the shower maximum parameter (which excludes distant muons which can mimick low energy gamma rays) to the one based mostly on Hillas parameters.

\subsection{Relative light scale between both telescopes}\label{sec:lightscale}
For observations at low zenith angles the density of Cherenkov light photons on the ground produced by a VHE $\gamma-$ray shower depends mostly on its energy and its impact parameter.
Except for a small dependence on the relative position of the shower axis with respect to the Geomagnetic field, due to the geomagnetic field effect (mostly pronounced at lowest energies, see e.g. \citealp{magic_gf}), the density is radially symmetric. 
Thus, it is possible to compare the light scale of both telescopes by selecting $\gamma$-like events from data in which the reconstructed impact parameter is similar in both telescopes \citep{ho03}. 
In the case of hadronic background events, such a correlation is much weaker due to the strong internal fluctuations and poor estimation of the impact parameter.
In order to obtain a nearly pure $\gamma$-ray sample, we apply rather strict cuts $Hadronness<0.2$ and $\theta^2<0.01$.

\begin{figure}[pt]
\centering 
\includegraphics[width=0.49\textwidth]{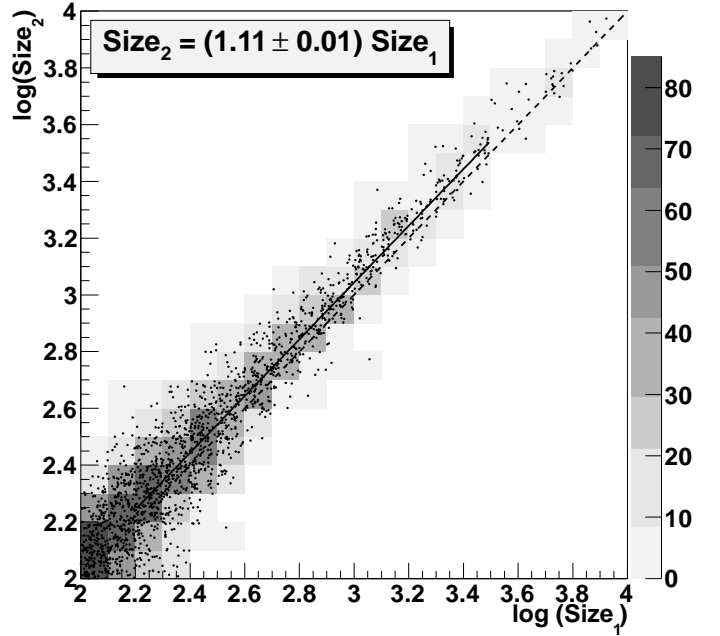}
\caption{
Correlation of $Size_2$ and $Size_1$ for $\gamma$-ray events obtained with Crab Nebula data. 
Only events with $Hadronness<0.2$, $\theta^2<0.01$, impact$_1<150$\,m, impact$_2<150$\,m and $|\mathrm{impact_1-impact_2}|<10$\,m are used. 
Individual events are marked by black dots, while the gray scale shows the total number of events in a given bin. 
The black solid line shows the result of the fit. 
The dashed line corresponds to $Size_1=Size_2$.
}\label{fig:sizescale}
\end{figure}
\begin{figure}[pt]
\centering 
\includegraphics[width=0.49\textwidth]{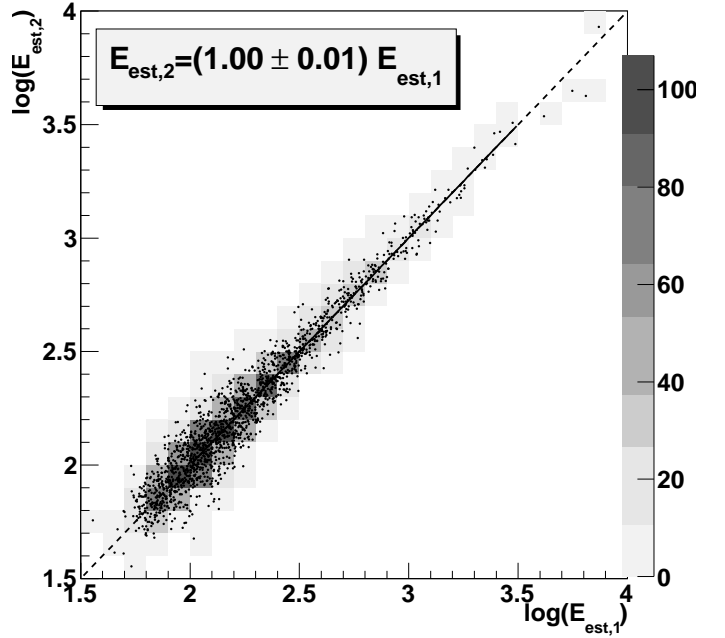}
\caption{
Correlation of the $\gamma$-ray estimated energy for  events (as in Fig.~\ref{fig:sizescale}) from the images recorded in each telescope separately. 
The black solid line shows the result of the fit. 
The dashed line corresponds to $E_{est,1}=E_{est,2}$.
}\label{fig:energyscale}
\end{figure}

The response of MAGIC-II is $(11 \pm 1_{stat})\%$ larger than that of MAGIC-I for showers observed at similar impact parameter (see Fig.~\ref{fig:sizescale}).
The result is a sum of multiple effects such as differences in the reflectivity of the mirrors, small differences between the two PMT populations, or uncertainity in the $F$-factor used for the calibration of each telescope.
The MC simulations are fine tuned to take into account this inter-telescope calibration. 
Therefore the estimated energy obtained independently from both telescopes is consistent (see Fig.~\ref{fig:energyscale}).

\subsection{Energy resolution}
\begin{figure}[tp]
\centering 
\includegraphics[width=0.49\textwidth]{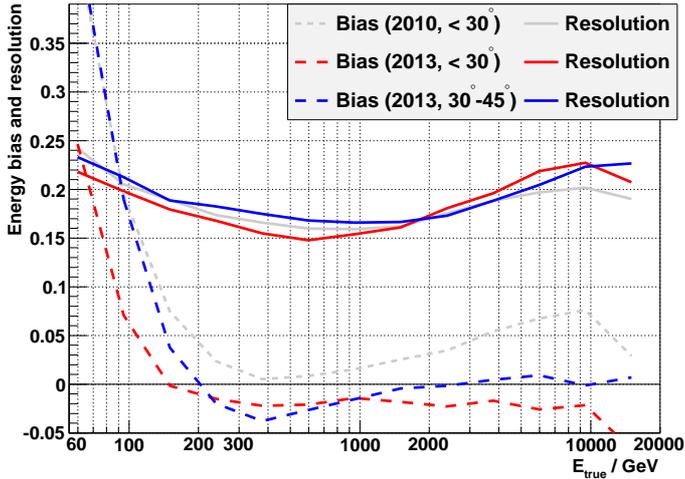}
\caption{
Energy resolution (solid lines) and bias (dashed lines) obtained from the MC simulations of $\gamma-$rays.
Events are weighted in order to represent a spectrum with a slope of $-2.6$. 
Red: low zenith angle, blue: medium zenith angle.
For comparison, pre-upgrade values from \cite{magic_stereo} are shown in gray lines.
}\label{fig:erec}
\end{figure}

We evaluate the performance of the energy reconstruction with $\gamma-$ray MC simulations. 
The simulations are divided into bins of true energy (5 bins per decade). 
In each bin we construct a distribution of $(E_{est}-E_{true})/E_{true}$ and fit it with a Gaussian function. 
The energy resolution is defined as the standard deviation obtained from this fit.
The bias of the energy reconstruction method can be computed as the mean value of the distribution.
The energy resolution and the bias of the MAGIC telescopes as a function of the true energy of the $\gamma$ rays are shown in Fig.~\ref{fig:erec}, and reported in Tables \ref{tab:enreslz} and \ref{tab:enresmz} for low and medium zenith angle respectively.

For low zenith angle observations in the energy range of a few hundred GeV the energy resolution falls down to about 15\%. 
For higher energies it degrades due to an increasing fraction of truncated images, and showers with high impact parameters as well as worse statistics in the training sample.  
Note that the energy resolution can be easily improved in the multi-TeV range with additional quality cuts (e.g. in the maximum reconstructed impact), however at the price of lowering the collection area.
At low energies the energy resolution is degraded, due to worse precision in the image reconstruction (in particular the impact parameters), and higher internal relative fluctuations of the shower. 
Above a few hundred GeV the absolute value of the bias is below a few percent. 
At low energies ($\lesssim100\,$GeV) the estimated energy bias rapidly increases due to the threshold effect. 
For observations at higher zenith angles the energy resolution is similar.
Since an event of the same energy observed at higher zenith angle will produce a smaller image, the energy resolution at the lowest energies is slightly worse. 
On the other hand, at multi-TeV energies, the showers observed at low zenith angle are often partially truncated at the edge of the camera, and may even saturate some of the pixels (if they produce signals of $\gtrsim750$\,phe in single pixels).
Therefore the energy resolution is slightly better for higher zenith angle observations. 
As the energy threshold shifts with increasing zenith angle, the energy bias at energies below 100 GeV is much stronger for higher zenith angle observations. 

The distribution $(E_{est}-E_{true})/E_{true}$ is well described by a Gaussian function in the central region, but not at the edges, where one can appreciate non-Gaussian tails. 
The energy resolution, determined as the sigma of the Gaussian fit, is not very sensitive to these tails. 
For comparison purposes, we also computed the RMS of the distribution (in the range $0 < E_{est} < 2.5 \cdot E_{true}$), which will naturally be sensitive to the tails of the $(E_{est}-E_{true})/E_{true}$. 
The RMS values are reported in Tables \ref{tab:enreslz} and \ref{tab:enresmz} for the low and medium zenith angles respectively. 
While the sigma of the Gaussian fit is in the range 15\%-25\%, the RMS values lie in the range 20\%-30\%.

\begin{figure}[t]
\centering 
\includegraphics[width=0.49\textwidth]{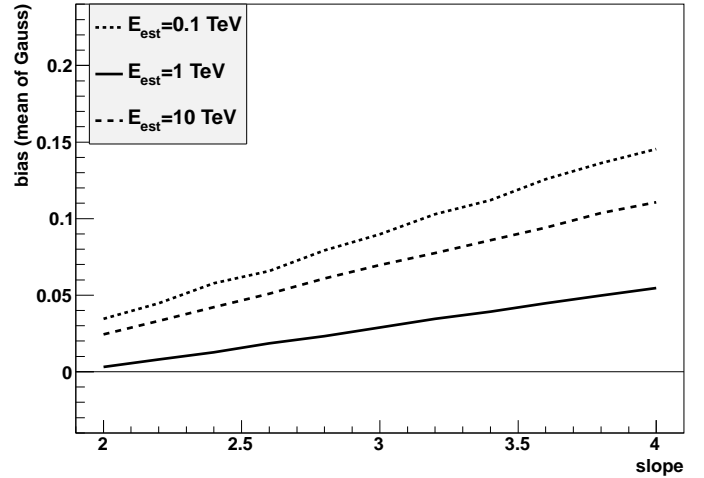}
\caption{
Energy bias as a function of the spectral slope for different estimated energies: $0.1\,$TeV (dotted line), $1\,$TeV (solid), $10\,$TeV (dashed).
Zenith angle below $30^\circ$.
}\label{fig:slopebias}
\end{figure}

When the data are binned according to estimated energy of individual events (note that, in contrary to MC simulations, in the data only the estimated energy is known) the value of the bias will change depending on the spectral shape of the source.
With steeper spectra more events will migrate from lower energies resulting in an overestimation of the energy. 
Note that this effect does not occur in the case of binning the events according to their true energy (as in Fig.~\ref{fig:erec}).
In Fig.~\ref{fig:slopebias} we show such a bias as a function of spectral slope for a few values of estimated energy. 
Note that the bias is corrected in the spectral analysis by means of an unfolding procedure \citep{magic_unfolding}. 

\begin{figure}[t]
\centering 
\includegraphics[width=0.49\textwidth]{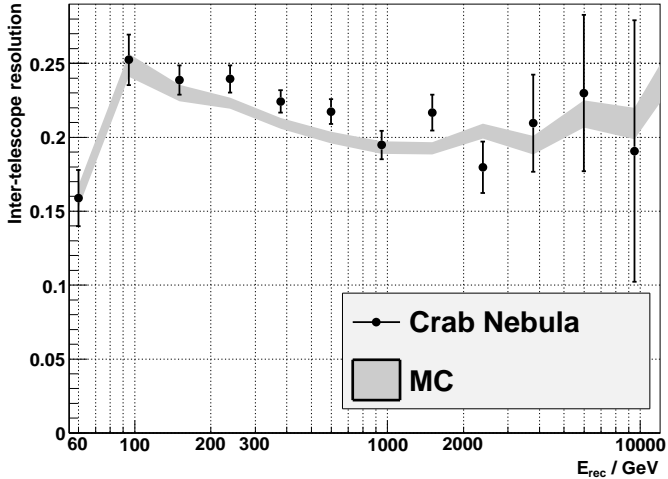}
\caption{
Standard deviation of the distribution of the relative difference between energy estimators from both telescopes as a function of the reconstructed energy for $\gamma$-ray MC (gray band) and Crab Nebula observations (black points). 
}\label{fig:erecrel}
\end{figure}

The energy resolution cannot be checked with the data in a straight-forward way and one has to rely on the values obtained from MC simulations. 
Nevertheless, we can use the fact of having two, nearly independent estimations of the energy, $E_{est,1}$ and $E_{est,2}$ from each of the telescopes to perform a consistency check.
We define relative energy difference as $RED = (E_{est,1}-E_{est,2})/E_{est}$. 
If the $E_{est,1}$ and $E_{est,2}$ estimators were completely independent the energy resolution would be $\approx RMS(RED)/\sqrt{2}$. 
In Fig.~\ref{fig:erecrel} we show a dependency of $RMS(RED)$ on the reconstructed energy.
The curve obtained from the data is consistent with the one of MC simulations within a few percent accuracy. 
The first point (between 45 and 75\,GeV) shows a sudden drop in $RMS(RED)$ compared to the other points, consistently in the data and MC simulations. 
Note that this point is below the analysis threshold, therefore it is mostly composed of peculiar events in which the shower produces more Cherenkov light than average for this energy. 
This results in a strong correlation of $E_{est,1}$ and $E_{est,2}$ allowing for a relatively low value of inter-telescope difference in estimated energy, and still a rather poor energy resolution.

\subsection{Spectrum of the Crab Nebula}\label{sec:sed}

\begin{figure}[pt]
\centering 
\includegraphics[width=0.49\textwidth]{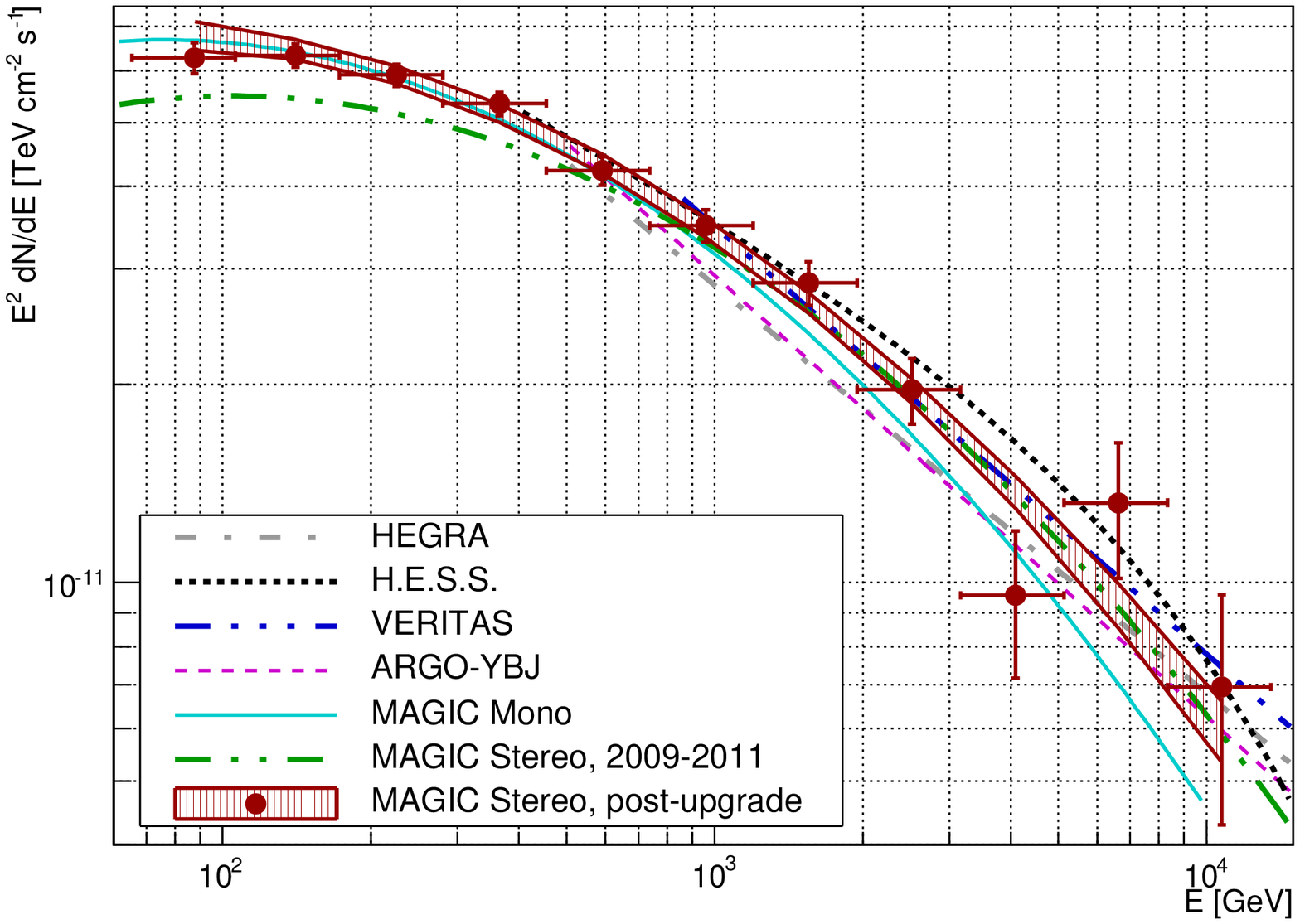}
\caption{
Spectral energy distribution of the Crab Nebula obtained with the MAGIC telescopes after the upgrade  (red points and shading) compared with other experiments: 
MAGIC-I (cyan solid, \citealp{magic_crab}), 
MAGIC Stereo, 2009-2011 (green dot-dot-dashed, \citealp{magic_rob_crab}), 
HEGRA (gray dot-dashed, \citealp{ah04}), 
VERITAS (blue triple dot dashed, \citealp{veritas_crab}),
ARGO-YBJ (magenta, dashed, \citealp{argo_crab}) 
and H.E.S.S. (black dotted, \citealp{ah06}). 
The vertical error bars show statistical uncertainties, while the horizontal ones represent the energy binning.  
}\label{fig:spectrum}
\end{figure}

In Fig.~\ref{fig:spectrum} we show the spectrum of the Crab Nebula obtained with the total (low + medium zenith angle) sample.
For clarity, the spectrum is presented in the form of spectral energy distribution, i.e. $E^2 \mathrm{d}N/\mathrm{d}E$.
In order to minimize the systematic uncertainty we apply $Hadronness$ and $\theta^2$ cuts with high $\gamma$-ray efficiency (90\% and 75\% respectively) for the spectral reconstruction.
The spectrum in the energy range 65$\,$GeV -- 13.5$\,$TeV can be fitted with a curved power-law:
\begin{equation}
\frac{\mathrm{d}N}{\mathrm{d}E}=f_0( E/ 1\ \mathrm{TeV}\,) ^{a+b\,\mathrm{\log}_{10}( E/ 1\,\mathrm{TeV}\,)}\ 
\mathrm{ [cm^{-2} s^{-1} TeV^{-1}]}. \label{eq:spectrum}
\end{equation} 
The parameters of the fit are:
$f_0 =( 3.39\pm 0.09_{\mathrm{stat}}) \times 10^{-11}$, 
$a=-2.51\pm 0.02_{\mathrm{stat}}$, 
and $b=-0.21\pm 0.03_{\mathrm{stat}}$.
The parameters of the spectral fit were obtained using the robust forward unfolding method which does not require regularization.
The forward unfolding requires however an assumption on the spectral shape of the source, and is insensitive to spectral features.
Therefore the individual spectral points were computed using the Bertero unfolding method \citep{magic_unfolding}.
The fit parameters obtained from both unfolding methods are consistent.

The spectrum obtained by MAGIC after the upgrade is consistent within $\sim25\%$ with the previous measurements of the Crab Nebula performed with other IACTs and earlier phases of the MAGIC telescopes.

\subsection{Angular resolution}\label{sec:angres}
Following the approach in \cite{magic_stereo}, we investigate the angular resolution of the MAGIC telescopes using two commonly used methods. 
In the first approach we define the angular resolution $\Theta_{Gaus}$ as the standard deviation of a 2-dimensional Gaussian fitted to the distribution of the reconstructed event directions of the $\gamma$-ray excess. 
Such a 2-dimensional Gaussian in the $\theta_x$ and $\theta_y$ space will correspond to an exponential fitting function for $\theta^2$ distribution.
The fit is performed in a narrow range, $\theta^2<0.025{[^\circ}^2]$, which is a factor $\sim2.5$ larger than the typical signal extraction cut applied at medium energies. 
Therefore it is a good performance quantity for looking for small extensions (comparable with angular resolution) in VHE $\gamma-$ray sources. 
In the second method we compute an angular distance, $\Theta_{0.68}$, around the source, which encircles 68\% of the excess events.
This method is more sensitive to long tails in the distribution of reconstructed directions. 
Note that while both numbers assess the angular resolution of the MAGIC telescopes, their absolute values are different, normally $\Theta_{Gaus}<\Theta_{0.68}$. 
For a purely Gaussian distribution $\Theta_{Gaus}$ would correspond to only 39\% containment radius of $\gamma$-rays originating from a point like source and $\Theta_{0.68}\approx1.5\,\Theta_{Gaus}$.

We use the low and medium zenith angle samples of the Crab Nebula to investigate the angular resolution.
Since the Crab Nebula is a nearby galactic source, it might in principle have an intrinsic size which would artificially degrade the angular resolution measured in this way. 
However, the extension of the Crab Nebula in VHE $\gamma$-rays was constrained to below $0.025^\circ$ \citep{hegra_crab}, making it a point-like source for MAGIC. 

\begin{figure*}[pt]
\centering 
\includegraphics[width=0.49\textwidth]{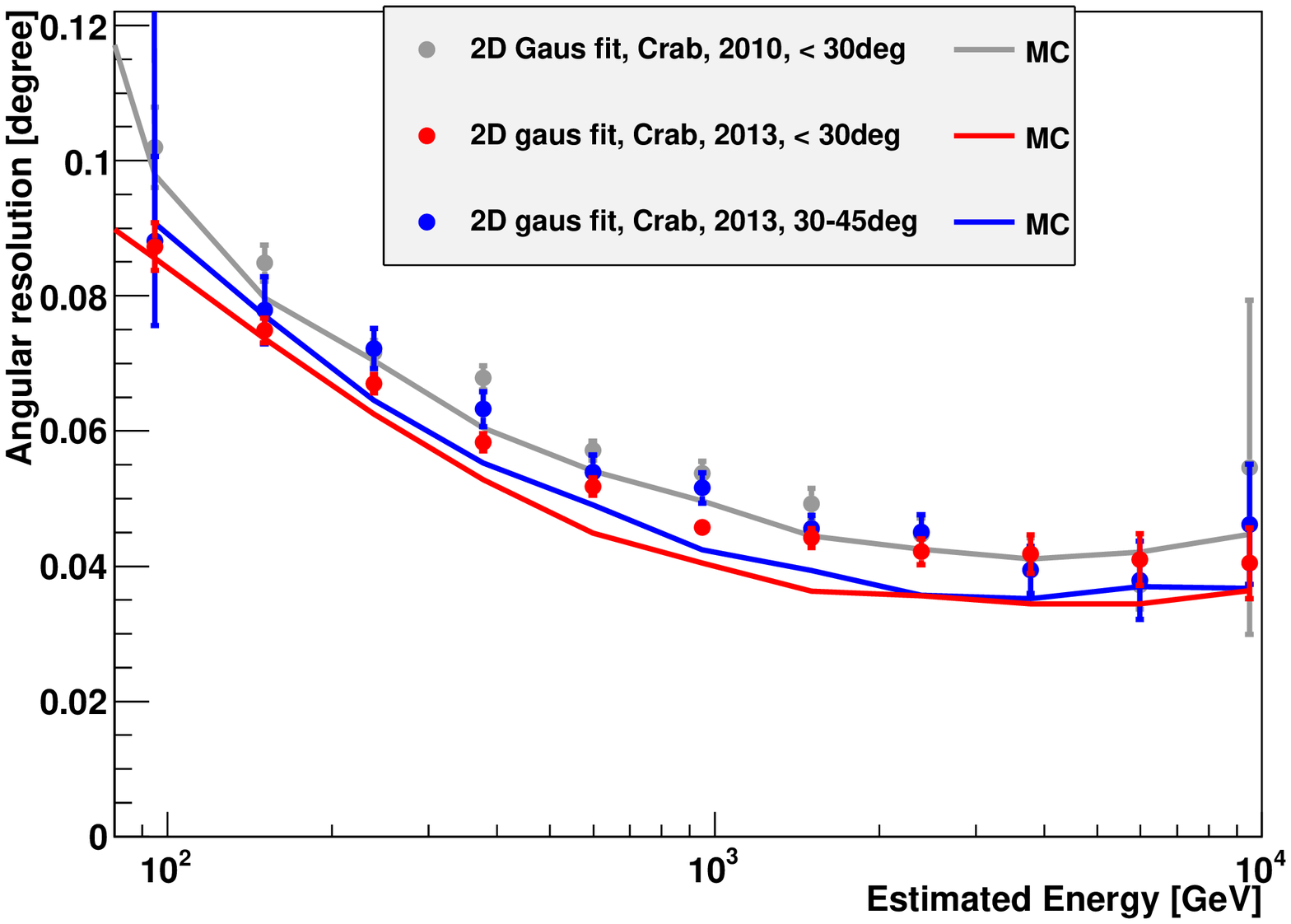}
\includegraphics[width=0.49\textwidth]{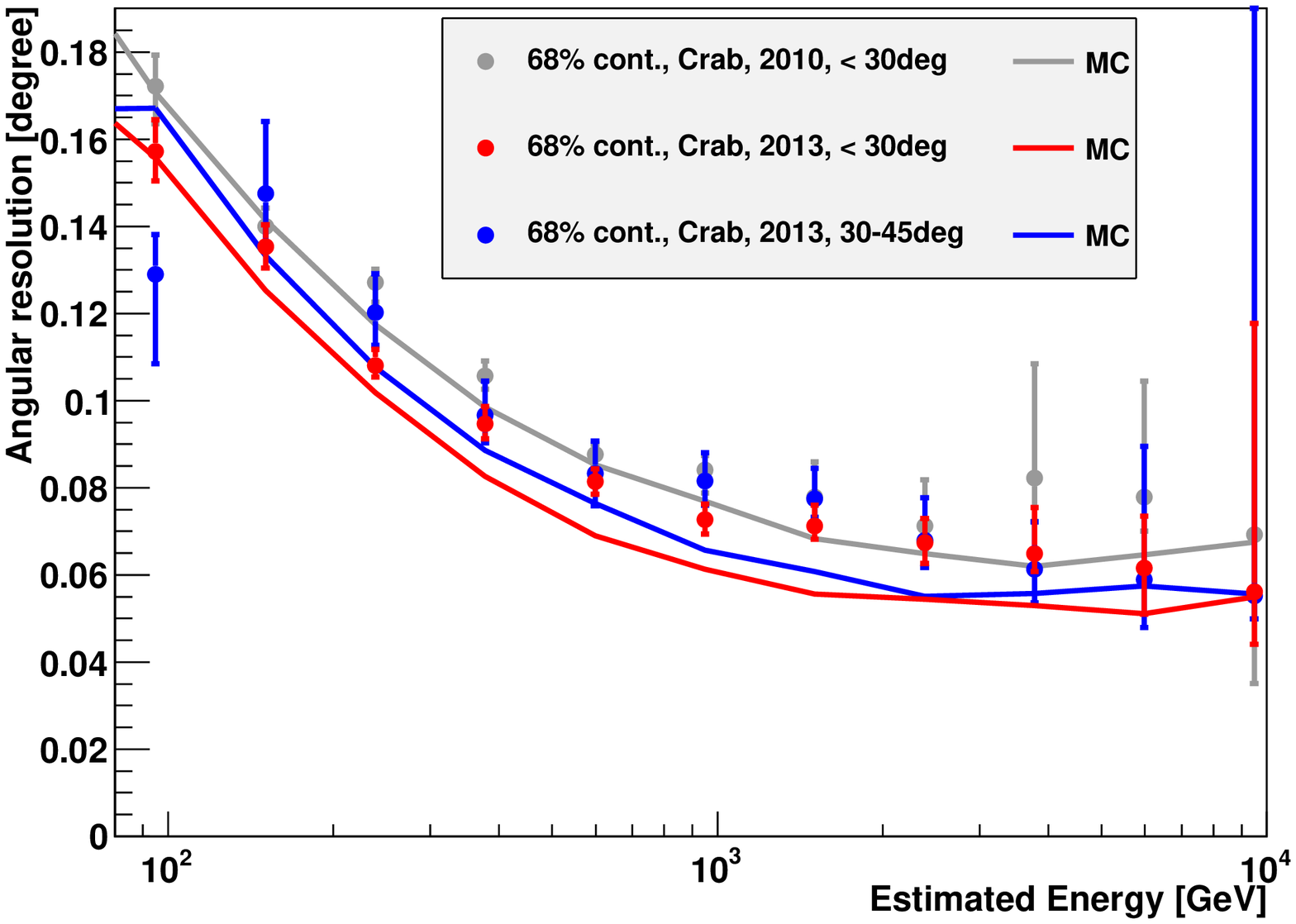}
\caption{
Angular resolution of the MAGIC telescopes after the upgrade as a function of the estimated energy obtained with the Crab Nebula data sample (points) and MC simulations (solid lines). 
Left panel: 2D Gaussian fit, right panel: 68\% containment radius.
Red points: low zenith angle sample, blue points: medium zenith angle sample.
For comparison the low zenith angle pre-upgrade angular resolution is shown as gray points \cite{magic_stereo}.
}\label{fig:angres}
\end{figure*}

The angular resolution obtained with both methods is shown in Fig.~\ref{fig:angres} and summarized in Table~\ref{tab:angres}.
At $250\,$GeV the angular resolution (from a 2D Gaussian fit) is $0.07^\circ$.
It improves with energy, as larger images are better reconstructed, reaching a plateau of $\sim 0.04^\circ$ above a few TeV. 
The angular resolution improved by about 5-10\% after the upgrade.
The improvement in angular resolution makes slightly more pronounced the small difference between the angular resolution obtained with MC simulations and the Crab Nebula data, also present in the pre-upgrade data.
The difference of $\sim 10-15\%$ is visible at higher energies and corresponds to an additional $0.02^\circ$ systematic random component (i.e. added in quadrature) between the MC and the data.

\begin{figure}[t]
\centering 
\includegraphics[width=0.49\textwidth]{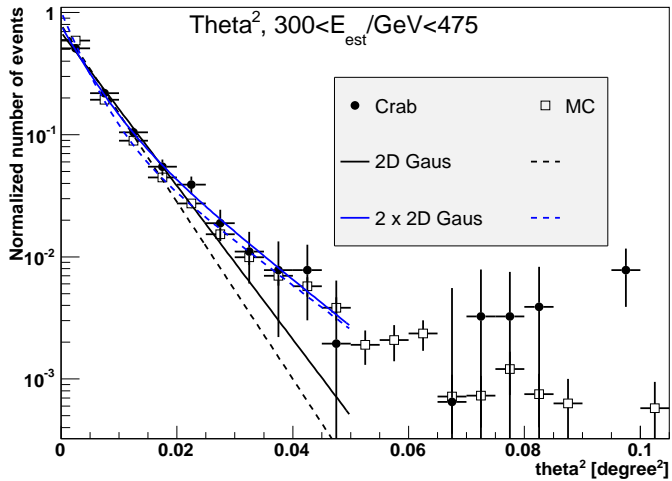}
\caption{
$\theta^{2}$ distribution of excess events for the Crab Nebula (filled circles, solid lines) and MC (empty squares, dashed lines) samples in the energy range of $300-475\,$GeV. 
The distributions are fitted with a single or a double two dimensional Gaussian (black and blue lines respectively).
}\label{fig:th2fit}
\end{figure}

The distributions of angular distances between the true and reconstructed source position can be reasonably well fitted with a single Gaussian for $\theta^2<0.025{[^\circ}^2]$.
Nevertheless, a proper description of the tail in the $\theta^2$ distribution requires a more complicated function (see Fig.~\ref{fig:th2fit}).
One possibility is to use a combination of two two-dimensional Gaussian distributions as in \cite{magic_stereo}. 
For example, the two-dimensional double Gaussian fit to the distribution shown in Fig.~\ref{fig:th2fit} for the Crab Nebula data yields $\chi^2/\mathrm{N_{dof}} = 1.5/6$ corresponding to a probability of $96.1\%$.

The tails of the PSF distribution do not have any practical impact on the background estimation. 
In the worst case scenario, which corresponds to observations close to the energy threshold and using three symmetrically reflected background regions, the contamination produced by the tails of the PSF is below 0.5\% of the signal excess, and hence negligible in comparison to other systematic uncertainties.

\begin{figure}[t]
\centering 
\includegraphics[width=0.49\textwidth]{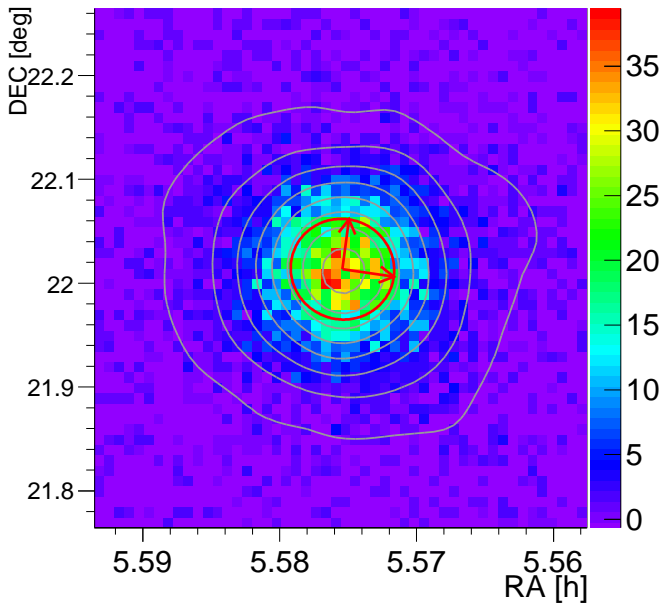}
\caption{
Two dimensional distribution of the excess events above 220 GeV from the Crab Nebula (color scale).
The significance contours (light gray lines) overlaid on the plot start with 5$\sigma$ for the most outer line with a step of 13$\sigma$ between neighboring lines. 
The distribution can be analytically fit by a 2D-Gaussian with RMS parameters in the two orthogonal directions reported by the red ellipse and the two arrows.
}\label{fig:crab2dpsf}
\end{figure}

Since MAGIC is a system of only two telescopes one may also expect some rotational asymmetry in the PSF shape due to a preferred axis connecting the two telescopes.
Note however, that MAGIC employs the DISP RF method for the estimation of the arrival direction, which is less affected by parallel images.
Therefore it is expected that the PSF asymmetry due to this effect will be reduced.  
In Fig.~\ref{fig:crab2dpsf} we present the distribution of excess events in sky coordinates obtained from the Crab Nebula. 
By computing the second order moments of the distribution and the x-y correlation we can derive the two perpendicular axes in which the spread of the distribution is maximal and minimal.
This is equivalent to perform a robust analytical fit with a two-dimensional Gaussian distribution. 
We find that the asymmetry of the PSF between these two axes is of the order of 10\%.
This asymmetry can be due to a mixture of effect such as optical coma aberration, having a preferred axis in the two telescope system and possibly a slightly different short term pointing precision in azimuth and zenith direction.

\subsection{Sensitivity}\label{sec:sens}
In order to provide a fast reference and comparison with other experiments we calculate the sensitivity of the MAGIC telescopes following the two commonly used definitions. 
For a weak source, the significance of an excess of $N_{\rm excess}$ events over a perfectly-well known background of $N_{\rm bkg}$ events can be computed with the simplified formula $N_{\rm excess}/\sqrt{N_{\rm bkg}}$.
Therefore, one defines the sensitivity $S_{Nex/\!\sqrt{Nbkg}}$ as the flux of a source giving $N_{\rm excess}/\sqrt{N_{\rm bkg}}=5$ after 50$\,$h of effective observation time.
The sensitivity can also be calculated using the \citet{lm83}, eq.~17 formula, which is the standard method in the VHE $\gamma$-ray astronomy for the calculation of the significances. 
Note that the sensitivity computed according to the \citeauthor{lm83} formula will depend on the number of OFF positions used for background estimation.

For a more realistic estimation of the sensitivity (in both methods), we apply conditions $N_{\rm excess}>10$ and $N_{\rm excess}> 0.05 N_{\rm bkg}$.
The first condition assures that the Poissonian statistics of the number of events can be approximated by a Gaussian distribution.
The second condition protects against small systematic discrepancies between the ON and OFF distributions, which may mimic a statistically significant signal if the residual background rate is large. 

\begin{figure}[t!]
\centering 
\includegraphics[width=0.49\textwidth]{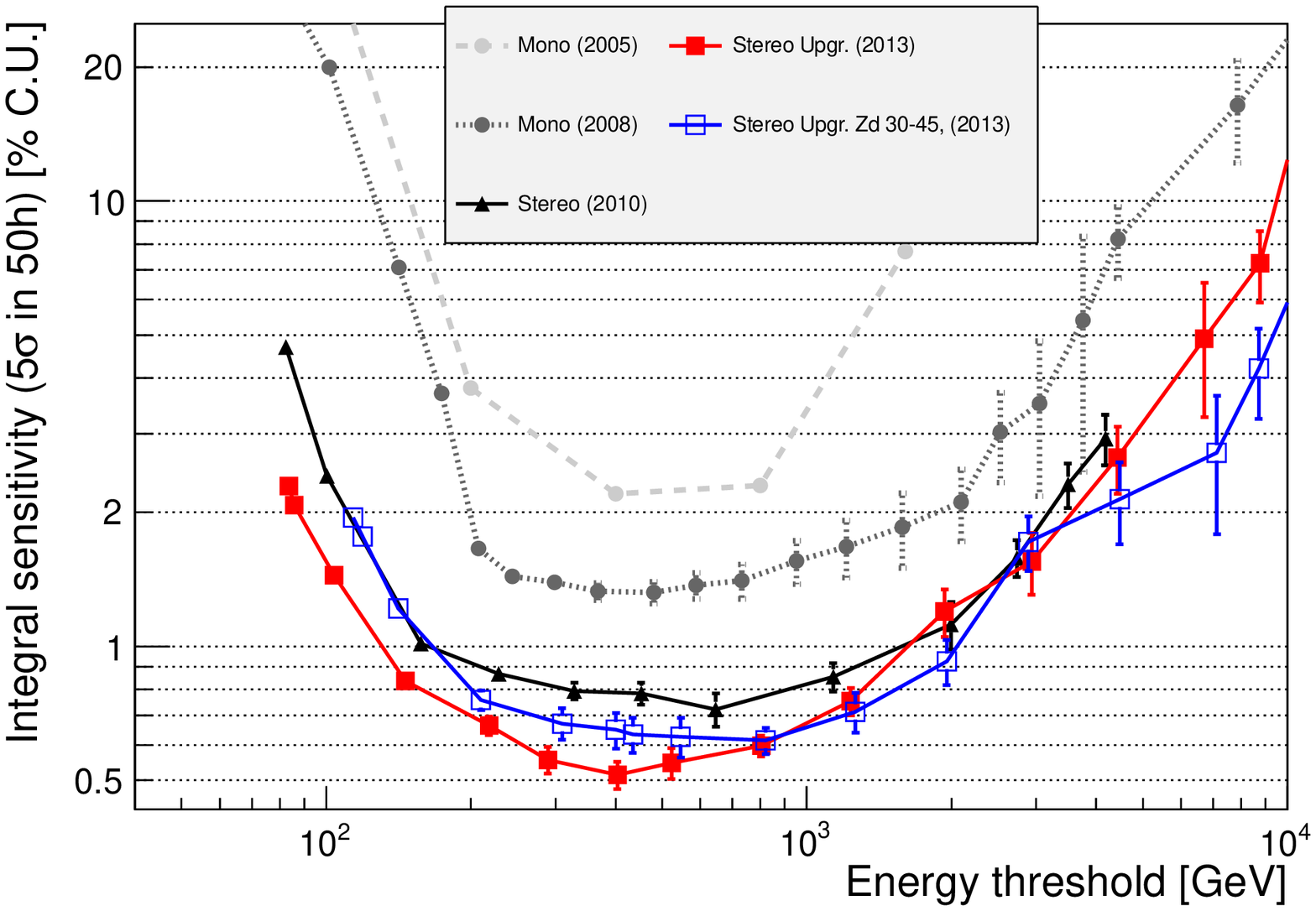}
\caption{
Evolution of integral sensitivity of the MAGIC telescopes, i.e. the integrated flux of a source above a given energy for which $N_{\rm excess}/\sqrt{N_{\rm bkg}}=5$ after $50\,\mathrm{h}$ of effective observation time, requiring $N_{\rm excess}>10$ and $N_{\rm excess}> 0.05 N_{\rm bkg}$.
Gray circles: sensitivity of the MAGIC-I single telescope with the Siegen (light gray, long dashed, \citet{magic_crab}) and MUX readouts (dark gray, short dashed, \citet{magic_stereo}).
Black triangles: stereo before the upgrade \citep{magic_stereo}.
Squares: stereo after the upgrade: zenith angle below $30^\circ$ (red, filled), $30-45^\circ$ (blue, empty)
For better visibility the data points are joined with broken lines.
}\label{fig:sens}
\end{figure}

The integral sensitivity of the different phases of the MAGIC experiment for a source with a Crab Nebula-like spectrum are shown in Fig.~\ref{fig:sens}. 
The sensitivity values both in Crab Nebula Units (C.U.) and in absolute units (following Eq.~\ref{eq:spectrum}) are summarized in Table~\ref{tab:intsens} for low zenith and in Table~\ref{tab:intsens2} for medium zenith angles.
We used here the $N_{\rm excess}/\sqrt{N_{\rm bkg}}=5$ definition, recomputing the original MAGIC-I mono sensitivities to include also the $N_{\rm excess}>10$ and $N_{\rm excess}> 0.05 N_{\rm bkg}$ conditions 
\footnote{Note that one of the main disadvantages of the mono observations was the very poor signal-to-background ratio at low energies, leading to dramatic worsening of the sensitivity. 
Using optimized cuts one can recover some of the sensitivity lost at the lowest energies for mono observations.}.

In order to find the optimal cut values in $Hadronness$ and $\theta^2$ in an unbiased way, we used an independent \emph{training} sample of Crab Nebula data. 
The size of the \emph{training} sample is similar to the size of the \emph{test} sample from which the final sensitivity is computed.
Different energy thresholds are achieved by varying a cut in the total number of photoelectrons of the images (for points $<300\,$GeV) or in the estimated energy of the events (above 300\,GeV).
For each energy threshold we perform a scan of cuts on the \emph{training} subsample, and apply the best cuts (i.e. those providing the best sensitivity on the \emph{training} subsample according to $N_{\rm excess}/\sqrt{N_{\rm bkg}}$ definition) to the main sample obtaining the sensitivity value.
The threshold itself is estimated as the peak of true energy distribution of MC events with a $-2.6$ spectral slope to which the same cuts were applied.

\begin{figure}[t]
\centering 
\includegraphics[width=0.49\textwidth]{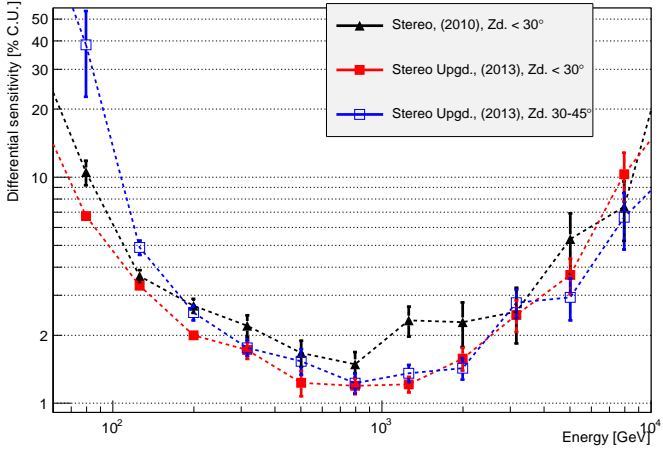}
\caption{
Differential (5 bins per decade in energy) sensitivity of the MAGIC Stereo system.
We compute the flux of the source in a given energy range for which $N_{\rm excess}/\sqrt{N_{\rm bkg}}=5$ with $N_{\rm excess}>10$, $N_{\rm excess}> 0.05 N_{\rm bkg}$ after 50$\,$h of effective time.
For better visibility the data points are joined with broken dotted lines.
}\label{fig:diffsens}
\end{figure}

The integral sensitivity evaluated above is valid only for sources with a Crab Nebula-like spectrum. 
To assess the performance of the MAGIC telescopes for sources with an arbitrary spectral shape, we compute the differential sensitivity. 
Following the commonly used definition,  we calculate the sensitivity in narrow bins of energy (5 bins per decade).
The differential sensitivity is plotted for low and medium zenith angles in Fig.~\ref{fig:diffsens}, and the values are summarized in Table~\ref{tab:diffsens} and \ref{tab:diffsens2} respectively.

The upgrade of the MAGIC-I camera and readout of the MAGIC telescopes has lead to a significant improvement in sensitivity in the whole investigated energy range.
The integral sensitivity reaches down to about 0.55\% of C.U. around a few hundred GeV in 50h of observations.
The improvement in the performance is especially evident at the lowest energies. 
In particular, in the energy bin 60-100 GeV, the differential sensitivity decreased from 10.5\% C.U. to 6.7\% C.U. reducing the needed observation time by a factor of 2.5.
Observations at medium zenith angle have naturally higher energy threshold.
Therefore the performance at the lowest energies is marred. 
Some of the sources, those with declination $>58^\circ$, or $<-2^\circ$ can only be observed by MAGIC at medium or high zenith angles. 
Sources with declination between $-2^\circ$ and $58^\circ$, can be observed either at low zenith angles, or at medium zenith angle with a boost in sensitivity at TeV energies at the cost of a higher energy threshold. 

\begin{figure}[t]
\centering 
\includegraphics[width=0.49\textwidth]{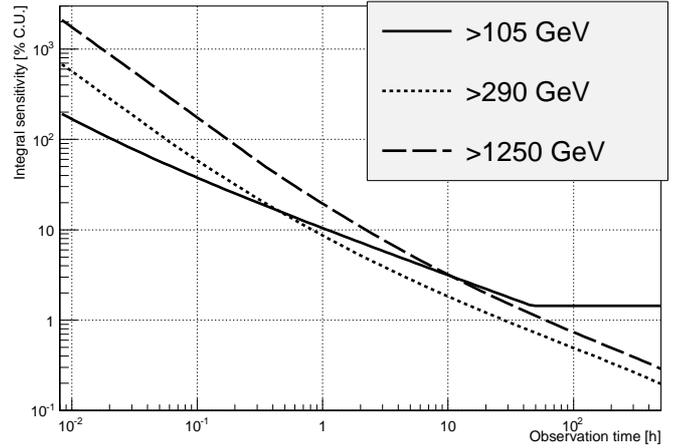}
\caption{
Dependence of the integral sensitivity of the MAGIC telescopes (computed according to $S_{\rm Li\&Ma, 3Off}$ prescription, see text for details) on the observation time, obtained with the low zenith angle Crab Nebula sample.
Different line styles show different energy thresholds: $>105$\,GeV (solid), $>290$\,GeV (dotted), $>1250$\,GeV (dashed).
}\label{fig:senstime}
\end{figure}

The sensitivity of IACTs clearly depends on the observation time which can be spent observing a given source. 
In particular for transient sources, such as gamma-ray bursts or flares from Active Galactic Nuclei, it is not feasible to collect 50h of data within the duration of such an event. 
On the other hand, long, multi-year campaigns allow to gather of the order of hundreds of hours (see e.g. $\sim$ 140h observations of M82 by VERITAS, \cite{veritas_m82}, $\sim$ 160h observations of Segue by MAGIC, \cite{magic_segue} or $\sim$ 180h NGC 253 by H.E.S.S., \cite{{ab12}}).
In Fig.~\ref{fig:senstime}, using the $\gamma$ and background rates from Table~\ref{tab:intsens}, we show how the sensitivity of the MAGIC telescopes depends on the observation time for different energy thresholds. 
For those exemplary calculations we use the Li\&Ma definition of sensitivity with typical value of 3 Off positions for background estimation. 
In the medium range of observation times the sensitivity follows the usual $\propto 1/\sqrt{time}$ dependence.
For very short observation times, especially for higher energies where the $\gamma$/hadron separation is very powerful, the limiting condition of at least 10 excess events leads to a dependence of $\propto 1/time$.
On the other hand, for very long observations the sensitivity saturates at low energies. 
Note that the observation time at which the sensitivity saturates might be shifted by using stronger cuts, offering better $\gamma$ to background rate, however at the price of increased threshold.

\subsection{Off-axis performance}
Most of the observations of the MAGIC telescopes are performed in the wobble mode with the source offset of $0.4^\circ$ from the camera center. 
However, in the case of micro-scans of extended sources with sizes much larger than the PSF of the MAGIC telescopes, a $\gamma$-ray signal might be found at different distances from the camera center \citep[see e.g.][]{magic_hess1857}.
Moreover, serendipitous sources (see e.g. detection of IC~310, \citealp{magic_ic310}) can occur in the FoV of MAGIC at an arbitrary angular offset from the pointing direction.
Therefore, we study the performance of the MAGIC telescopes at different offsets from the center of the FoV with dedicated observations of the Crab Nebula at non-standard wobble offsets (see Table~\ref{tab:data}).

\begin{figure}[t]
\centering 
\includegraphics[width=0.49\textwidth]{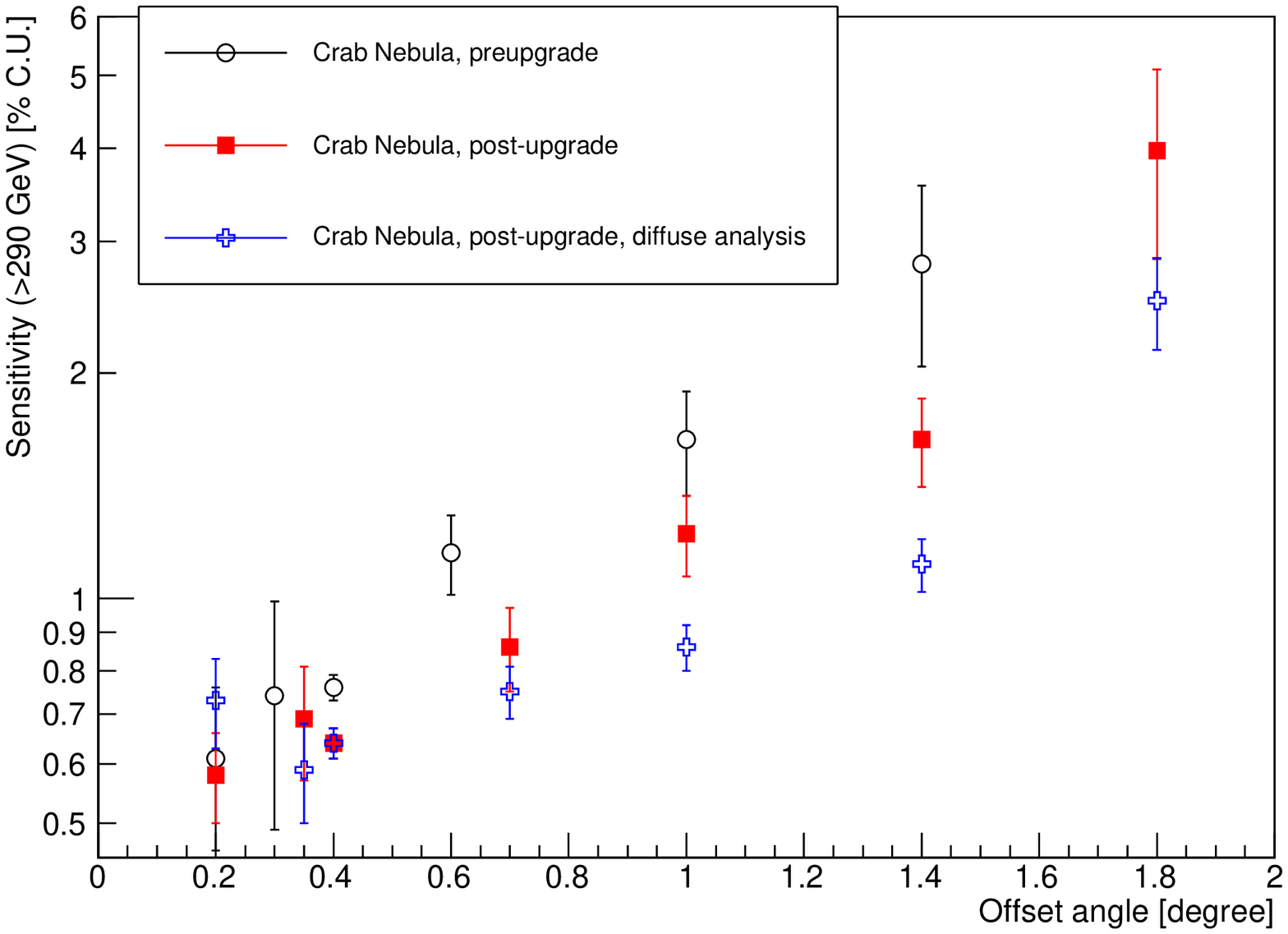}
\includegraphics[width=0.49\textwidth]{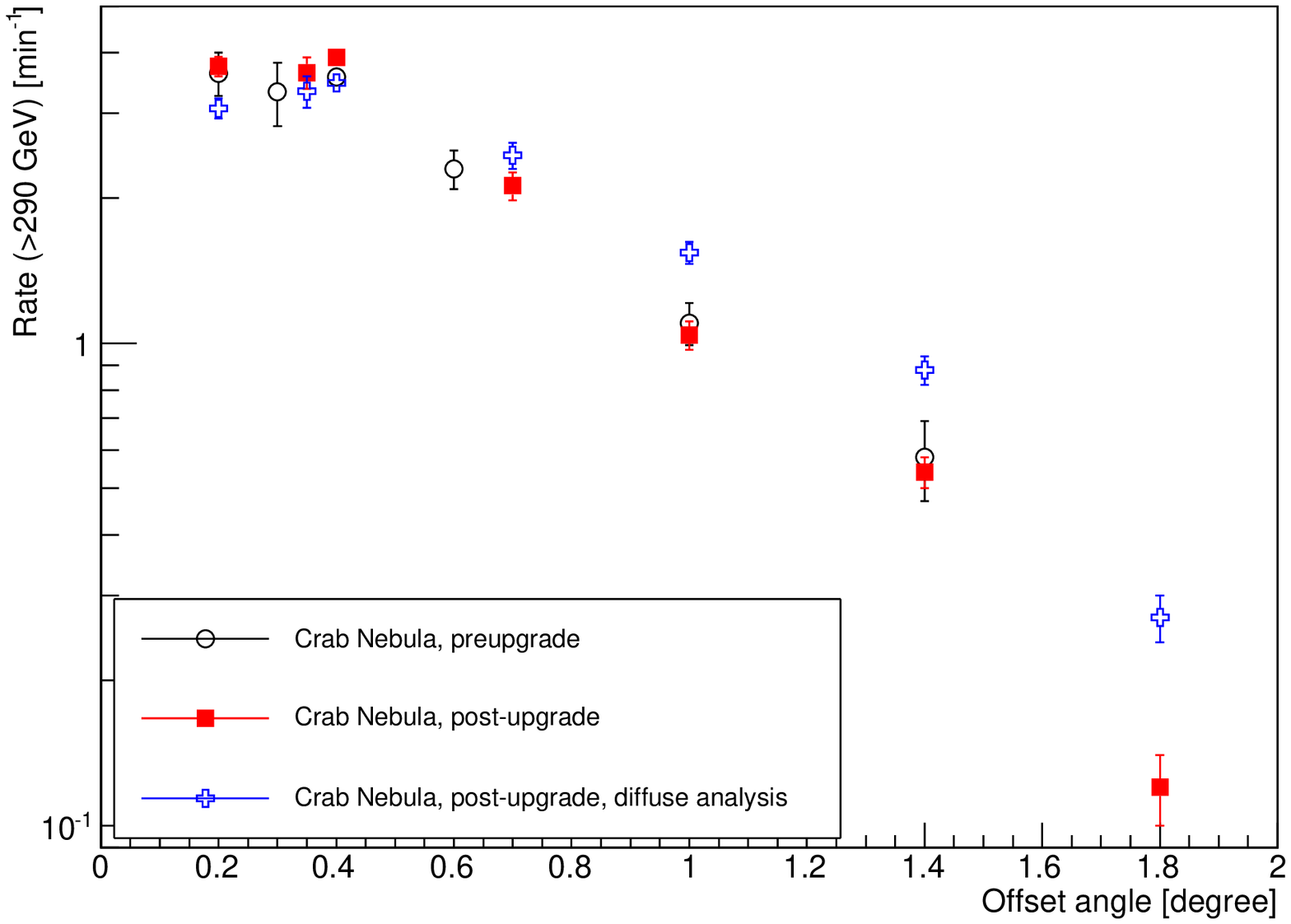}
\caption{
Top panel: integral sensitivity, computed according to $S_{Nex/\!\sqrt{Nbkg}}$ prescription (see Section~\ref{sec:sens}), above 290$\,$GeV for low zenith angle observations at different offsets, $\xi$, from the camera center. 
Bottom panel: corresponding (obtained with the same cuts as the sensitivity), $\gamma$-ray rates $R(\xi)$.
Black empty circles: data from before the upgrade \citep{magic_stereo}, 
red filled squares: current data (see Table~\ref{tab:data})
blue empty crosses: current data with ``diffuse'' analysis.
}\label{fig:offset}
\end{figure}

For easy comparison with the results presented in \cite{magic_stereo}, we first compute the integral sensitivities as a function of the wobble offset at the same energy threshold of $290\,$GeV.
We first apply the same kind of analysis as was used in \citealp{magic_stereo}, i.e. where the $\gamma$/hadron separation and direction reconstruction is trained with MC simulations generated at the standard offset of $\xi=0.4^\circ$ (see red filled squares in Fig.~\ref{fig:offset}).
The upgrade of the MAGIC telescopes has improved the off-axis performance. 
For example, the sensitivity at offsets of $\sim 1^\circ$ has improved by $\sim 25\%$, which is more than the global 15\% improvement seen at the usual offset of $\sim 0.4^\circ$.
Interestingly there is not much difference in the $\gamma$ rates associated with these sensitivity values.
This suggests that most of the improvement in sensitivity comes from a better image reconstruction, possibly thanks to the higher pixelization of the new MAGIC-I camera, rather than from triggering more events due to larger trigger region. 
We performed also a second analysis, the so-called ``diffuse'' one (see blue empty crosses in Fig.~\ref{fig:offset}).
In this case the $\gamma$/hadron separation and direction reconstruction is trained with MC simulations of $\gamma$-rays with a diffuse origin within a $1.5^\circ$ radius from the camera centre.
We find this analysis to provide a better performance at large offset angles.

Note that depending on the offset angle of the source different number of background estimation regions can be used, which will affect the significance computed according to the prescription of \citet{lm83}.  
For large offset values more than the standard 3 regions can be used. 
However as the uncertainty then is dominated by the fluctuations of the number of ON events, the significance saturates fast, and even in this case the extra gain does not exceed 10\%.

\subsection{Extended sources}
Some of the sources might have an intrinsic extension. 
The sensitivity for detection such sources is degraded for two reasons. 
First of all, the signal is diluted over a larger part of the sky.
This forces us to loosen the angular $\theta^2$ cut and hence accept more background.
The new cut value can be roughly estimated to be 
\begin{equation}
\theta_{cut} = \sqrt{\theta_{0}^2 + \theta_{s}^2}, \label{eq:thetacut}
\end{equation}
where $\theta_{0}$ is a cut for a point-like source analysis, and $\theta_{s}$ is a characteristic source size.
As the background events show a nearly flat distribution of $\mathrm{d}N/\mathrm{d}\theta^2$ such a cut will increase the background by $\theta_{cut}^2$.
The looser $\theta^2$ cut will affect the sensitivity in two ways. 
First of all, the sensitivity will be degraded by a factor of $\sqrt{\theta_{cut}^2}=\theta_{cut}$ due to accepting more background events.
Note however that the acceptance of the $\theta_{cut}$ cut for $\gamma$-rays can be larger than the acceptance of $\theta_{0}$ cut for a point like source, as the $\gamma$-rays originating from the center of the source will still be accepted even if they are strongly misreconstructed.

\begin{figure}[pt]
\centering 
\includegraphics[width=0.49\textwidth]{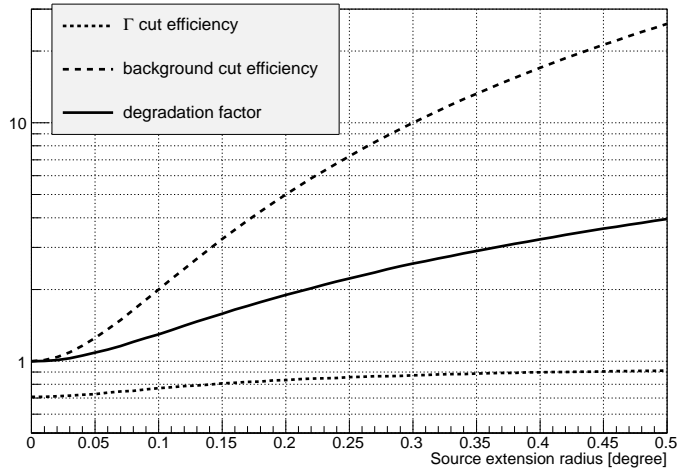}
\caption{
Dashed line: dependence of the amount of background integrated up to a cut determined by Eq.~\ref{eq:thetacut} as a function of the radius of the source, normalized to the background for a point like source.
Dotted line: fraction of the total $\gamma$ events contained within the cut.
Solid line: sensitivity for an extended source divided by sensitivity for a point like source.
A flat surface profile of the emission is assumed in the calculations.
}\label{fig:extended}
\end{figure}

For the further calculations let us assume $\theta_{0}=0.1$ (comparable to $\theta_{0.68}$ at the energies of a few hundred GeV, see Section~\ref{sec:angres}) and a PSF shape as described by Fig.~\ref{fig:th2fit} (note that the PSF is not much affected by the distance from the center of the camera, \cite{magic_stereo}) and a source with a flat surface brightness up to $\theta_{s}$. 
In Fig.~\ref{fig:extended} we show the acceptance for background and $\gamma$ events.
We also compute a sensitivity ``degradation factor'', defined as the square root of the background acceptance divided by the $\gamma$ acceptance and normalized to 1 for a point like source. 
As an example, let us assume a source with a radius of $0.5^\circ$. 
The optimal cut $\theta_{s}=0.51$ computed according to Eq.~\ref{eq:thetacut} results in 26 times larger background than with cut $\theta_{0}=0.1$. 
This would correspond to $\approx5$ times worse sensitivity, however the cut contains $\approx90\%$ of $\gamma$ events, significantly larger than $\approx 70\%$ efficiency for a point like cut. 
Therefore the sensitivity is degraded by a smaller factor, $\approx4$.

A second effect which can degrade the sensitivity for extended sources is the loss of collection area for higher offsets from the camera center. 
For a source radius of e.g. $0.5^\circ$, the $\gamma$-rays can be observed up to an offset of $0.9^\circ$ from the camera center. 
For such large offsets, the collection area is nearly a factor of 3 smaller than in the camera center. 
Using the $\gamma$-rates, which are proportional to the collection area,  shown in Fig.~\ref{fig:offset} we can compute the average rate of $\gamma$ rays for an arbitrary source profile.
For this example of a source with constant surface density and a radius of $0.5^\circ$ it turns out that the total average collection area is lower only by $\approx20\%$ than for a point like source at the usual wobble offset of $0.4^\circ$.
However, since a similar drop happens also for the background events, the net degradation of the sensitivity due to this effect is only $\sim10\%$. 

Finally, we compute the radius $\Xi$ of the MAGIC {\it effective field of view}.
It is defined such that observations of an isotropic gamma-ray flux with a hypothetical instrument with a flat-top acceptance $R'(\xi)= R(0)$ for $\xi<\Xi$, and $R'(\xi) = 0$ for $\xi>\Xi$, would yield the same number of detected gamma rays as with MAGIC, when no cuts on the arrival direction are applied. 
We can therefore obtain $\Xi$ from the condition $\int_{0}^{\Xi} 2\pi\,\xi\,R(0)\,d\xi = \int_{0}^{1.8^\circ}2\pi\,\xi\,R(\xi)\,d\xi$, where $R(\xi)$ is shown in bottom panel of Fig.~\ref{fig:offset}, yielding $\Xi=1^\circ$. 
We note, however, that standard observations of sources with an extension larger than 0.4$^\circ$ are technically difficult, as in that case the edge of the source would fall into the background estimation region. 
Nevertheless, the {\it effective field of view} is a useful quantity for non-standard observations of diffuse signals like, e.g.\ the cosmic electron flux \citep{magic_electrons, ah09}.

\section{Systematic uncertainties}\label{sec:systematics}

The systematic uncertainties of the IACT technique stem from many small individual factors which are only known with limited precision, and possibly change from one night to another. 
Most of those factors (e.g. uncertainities connected with the atmosphere, reflecitivity of the mirrors) were not affected by the upgrade and hence the values reported in \cite{magic_stereo} are still valid for them.
In this section we evaluate the component of the total systematic uncertainity of the MAGIC telescopes which changed for observations after the upgrade. 
We also estimate the total systematic uncertainty for various observation conditions.

\subsection{Background subtraction}
Dispersion in the PMT response (including also a small number of ``dead'' pixels) and NSB variations  (e.g. due to stars) across the field of view of the telescopes cause a small inhomogeneity in the distribution of the events in the camera plane.
In addition, stereoscopy with just two telescopes produces a natural inhomogeneity, with the distribution of events being slightly dependent on the position of the second telescope. 
This effect was especially noticeable before the upgrade, due to the smaller trigger area of the old MAGIC-I camera.
Both of these effects result in a slight rotational asymmetry of the camera acceptance. 
The effect is minimized by wobbling such that the source and background estimation positions in the camera are being swapped. 
Before the upgrade the systematic uncertainity of the background determination was $\lesssim2\%$ \citep{magic_stereo}. 
We performed a similar study on a data sample taken after the upgrade.
We compare the background estimated in two reflected regions on the sky, without known $\gamma$-ray sources.
In order to achieve the needed statistical accuracy we apply very loose cuts. 
In the lowest energy range we obtain $56428 \pm 238$ events in one position versus $55940 \pm 237$, i.e. a difference of $(0.9 \pm 0.6)\%$ consistent within the statistical uncertainty.
A similar study in the medium energy range results in $7202 \pm 85$ versus $7233 \pm 85$ events which are consistent within the statistical uncertainties: $(-0.4 \pm 1.7)\%$. 
We conclude that due to the larger trigger region this uncertainty has been reduced now to $\lesssim1\%$. 

Note that the effect of the background uncertainty depends on the signal to background ratio.
In case of a strong source, where the signal $\gtrsim$ background, it is negligible.
However for a very weak source, with e.g. a signal to background ratio of $\sim 5\%$ , the additional systematic uncertainty on the flux normalization just from the uncertainty of the background can amount up to $\sim 20\%$.
Moreover, as it will be energy dependent, it might lead to an additional uncertainty in the spectral index. 
Let us consider a hypothetical weak source with a spectrum reconstructed between $E_{\min}$ and $E_{\max}$. 
The signal to background ratio of this source is $\mathrm{SBR_{LE}}$ and $\mathrm{SBR_{HE}}$ in the energy range $E_{\min} - \sqrt{E_{\min} E_{\max}}$ and $\sqrt{E_{\min} E_{\max}} - E_{\max}$ respectively.
In this case we can roughly estimate the systematic uncertainty on the spectral index due to background inhomogeneity as:
\begin{equation}
\Delta\alpha_\mathrm{SBR} = 2\times \frac{\sqrt{(1\%/\mathrm{SBR_{LE}})^2 + (1\%/\mathrm{SBR_{HE}})^2}}{\ln(E_{\max} / E_{\min})}, \label{eq:bgdsyst}
\end{equation}
where the 1\% comes from the precision at which the background is estimated. 
Note that this formula was already used e.g. in \cite{magic_ngc1275, magic_pks1424}, however with a larger value of background uncertainty.
For a strong source observed in a broad energy range, Eq.~\ref{eq:bgdsyst} gives a negligible number, e.g. for source observed between 0.08-6\,TeV with a SBR of 25-60\% $\Delta\alpha_\mathrm{SBR}=0.02$.
However for a weak source, e.g. observed between 0.1 and 0.6\,TeV with SBR = 6-15\% we obtain $\Delta\alpha_\mathrm{SBR}=0.2$, increasing the total systematic uncertainty on the spectral index from $0.15$ to $0.25$.

\subsection{Pointing accuracy}
\begin{figure}[t]
\centering 
\includegraphics[width=0.49\textwidth]{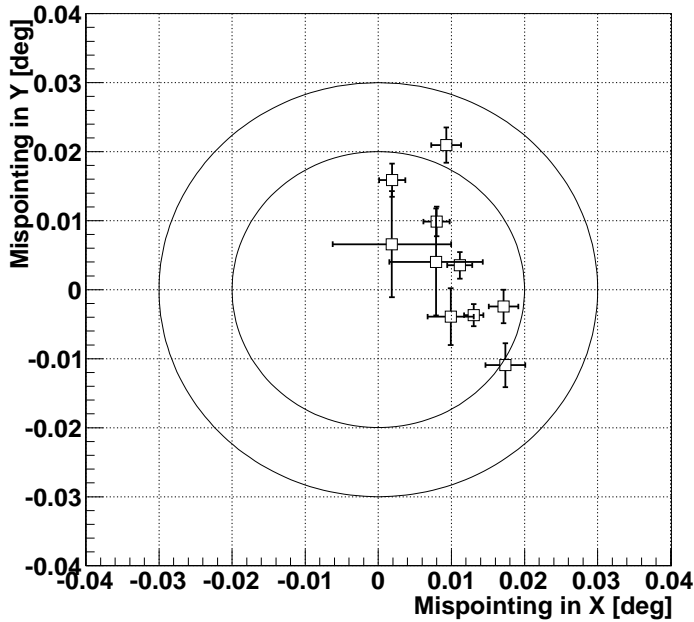}
\caption{
The difference in the reconstructed and nominal source position of the Crab Nebula during 10 nights of observations.
The two circles show a distance of $0.02^\circ$ and $0.03^\circ$. 
}\label{fig:mispointing}
\end{figure}

During the observations, the gravitational loads lead to a slight deformation of the telescopes structure and sagging of the camera. 
Most of the effect is corrected by the active mirror control and simultaneous observations of reference stars with CCD cameras placed at the center of the telescope's reflectors.
However, a slight residual mispointing of the telescopes can affect the precision with which the position of a $\gamma-$ray source can be obtained. 
In order to evaluate this effect we analyse the Crab Nebula data night by night.
For each night we construct a two dimensional distribution of the difference between the reconstructed and nominal source position in camera coordinates. 
After subtracting the background, the distribution is fitted with a two dimensional Gaussian to determine a possible systematic offset (see Fig.~\ref{fig:mispointing}). 
Note that the 'Y' direction in the camera correspond to the Zenith axis of the telescope.
We conclude that the systematic uncertainty on the reconstructed source position is $\lesssim 0.02^\circ$, comparable to value obtained in \cite{magic_halo}.

\subsection{Energy scale}
The absolute energy scale of IACTs is hard to determine. 
There are many systematic effects such as imprecise knowledge of the atmospheric transmission, mirror reflectivity, properties of the PMTs, etc. which affect it.
If determined solely from the best knowledge of these parameters, it is expected to be accurate within 15-17\% as shown in \cite{magic_stereo}.
The knowledge of the absolute energy scale of the MAGIC telescopes is validated by using inter-telescope calibration (see Section~\ref{sec:lightscale}) and from the analysis of muon rings \citep{iact_muons,magic_muon}.
Both of those methods are burdened by their own systematic uncertainities.
The inter-calibration of the telescopes using $\gamma$-ray events with similar impact parameters improves greatly the relative size scale, however has a very weak handle on the absolute light scale of both telescopes. 
On the other hand, the muon analysis suffers from the fact that the light seen from selected muon events is emitted mostly up to the height of $\sim$800\,m above the telescopes, while the light from $\gamma$-ray events is generated mainly at the height of $10\,$km.
This in addition makes the light spectrum of muon events shifted to lower wavelengths. 
The muon calibration used in MAGIC depends on comparison with muon MCs, which introduces additional systematic uncertainties due to possible data/MC mismatches at the shortest wavelengths.

A small miscalibration of the energy scale will affect the reconstructed spectrum in two ways, both having the highest impact at the lowest energies.
For example, let us assume that MC simulations have a higher light scale, i.e. for a given energy the amount of light generated by a $\gamma-$ray shower at a given energy predicted by the MC simulations is larger than for real showers.
In this case, real $\gamma-$ray showers will have a smaller \emph{Size} parameter than predicted by the MC simulations. 
Thus some of them may not survive a data quality size cut, or not even trigger the telescopes lowering the real collection area w.r.t. the one predicted from MC simulations.
This would artificially lower the reconstructed flux, especially at the low energies.
On the other hand, the overestimation of the light shed into the cameras by the shower will introduce an unrecoverable bias in the energy estimation migrating events to lower energies.
For the medium energies where the collection area is quite flat and the usual $\gamma$-ray spectra drop rapidly this will also artificially lower the reconstructed spectrum. 
For the lowest energies, below the analysis threshold, where the collection area drops very fast, more events will migrate in a given energy bin than escape from it, artificially increasing the reconstructed flux by a pile-up effect. 
It is not obvious to determine which of those effects will be dominant at a given energy.
It will depend on the precise shape of the collection area, the spectrum of the source and the total miscalibration of the light scale.
Therefore, the systematic error in the light scale can shift the spectrum at the lowest energies in both ways.

\begin{figure*}[pt]
%\centering 
\includegraphics[width=0.49\textwidth]{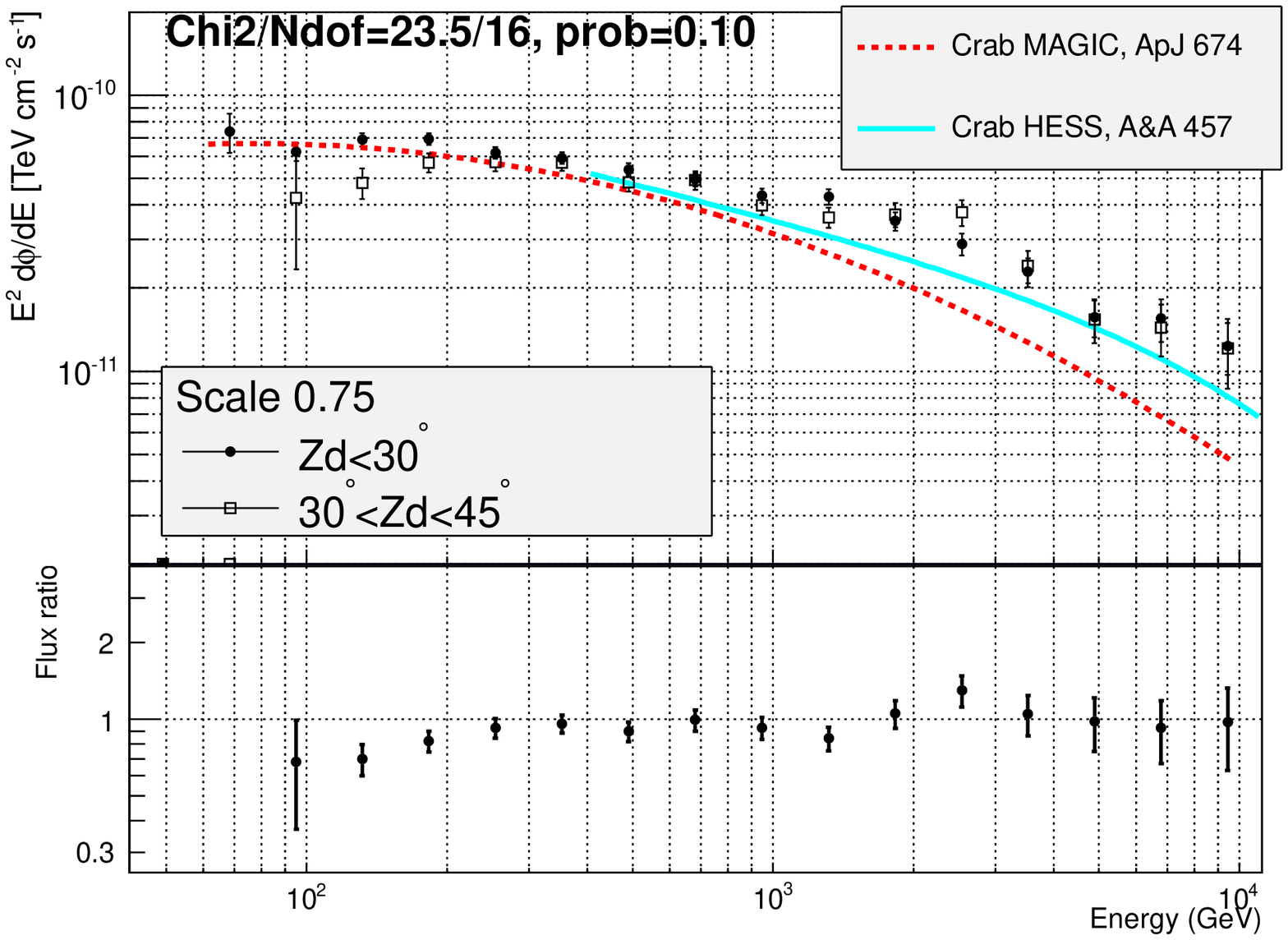}
\includegraphics[width=0.49\textwidth]{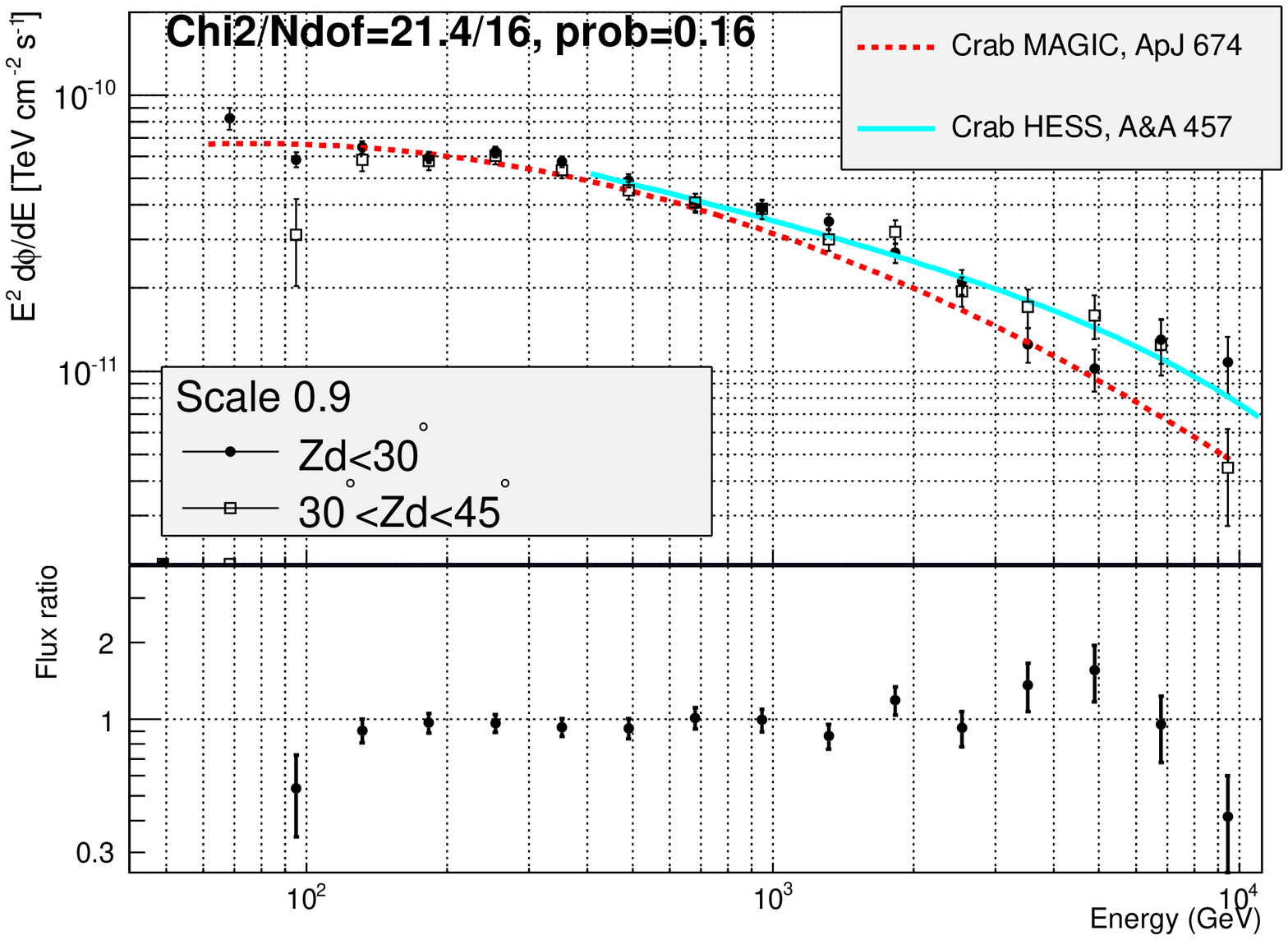} \\
\includegraphics[width=0.49\textwidth]{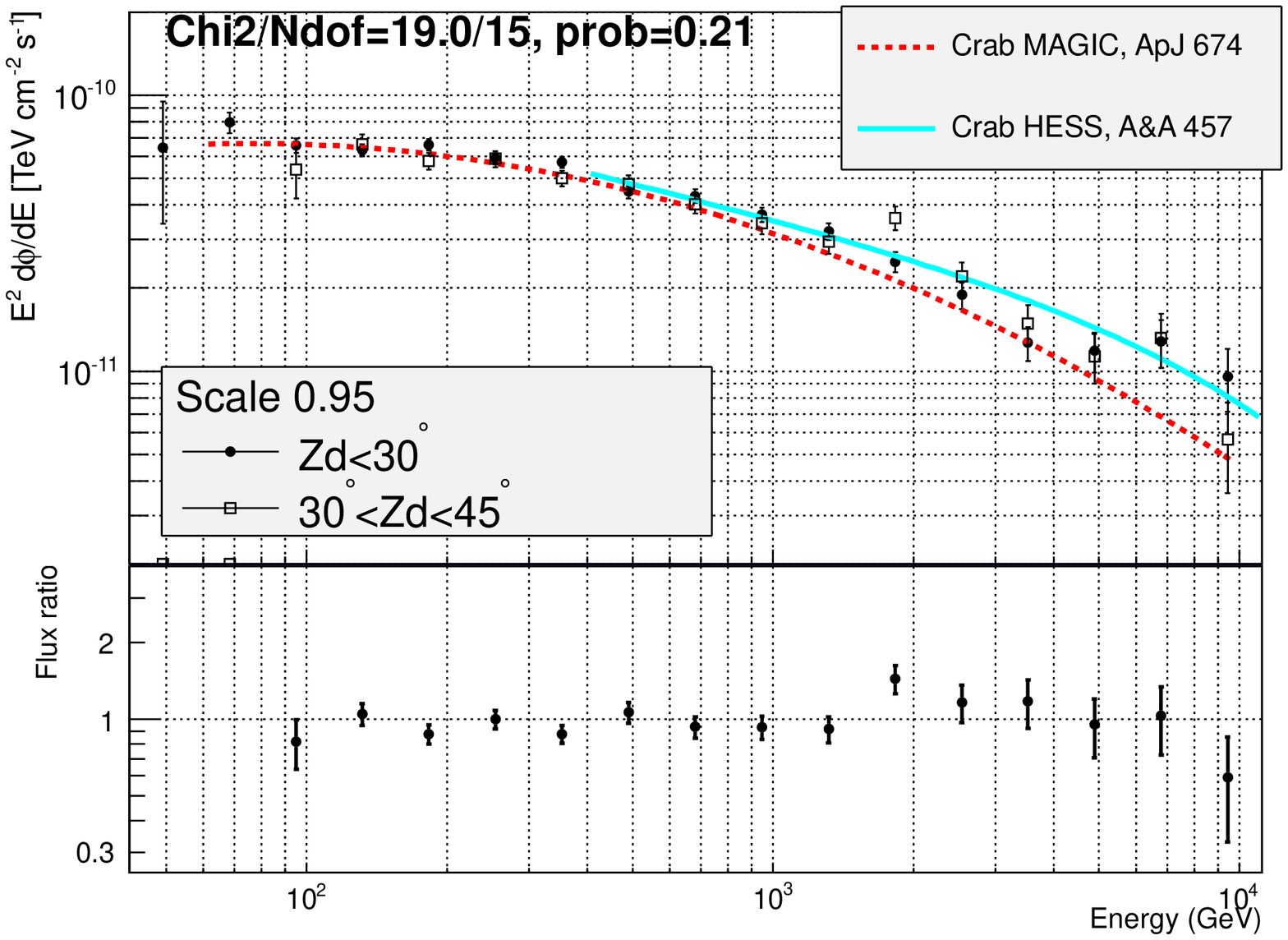} 
\includegraphics[width=0.49\textwidth]{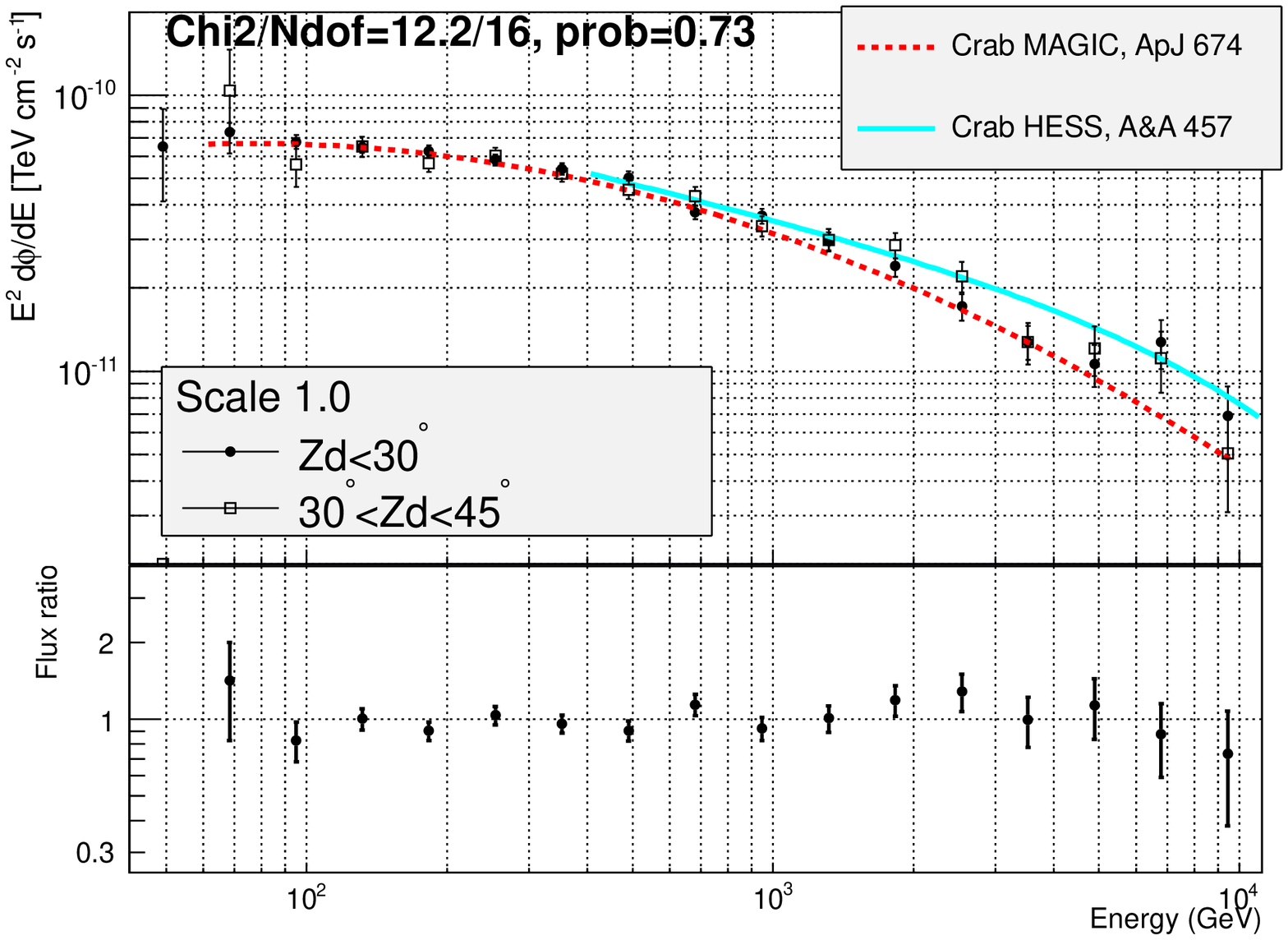} \\
\includegraphics[width=0.49\textwidth]{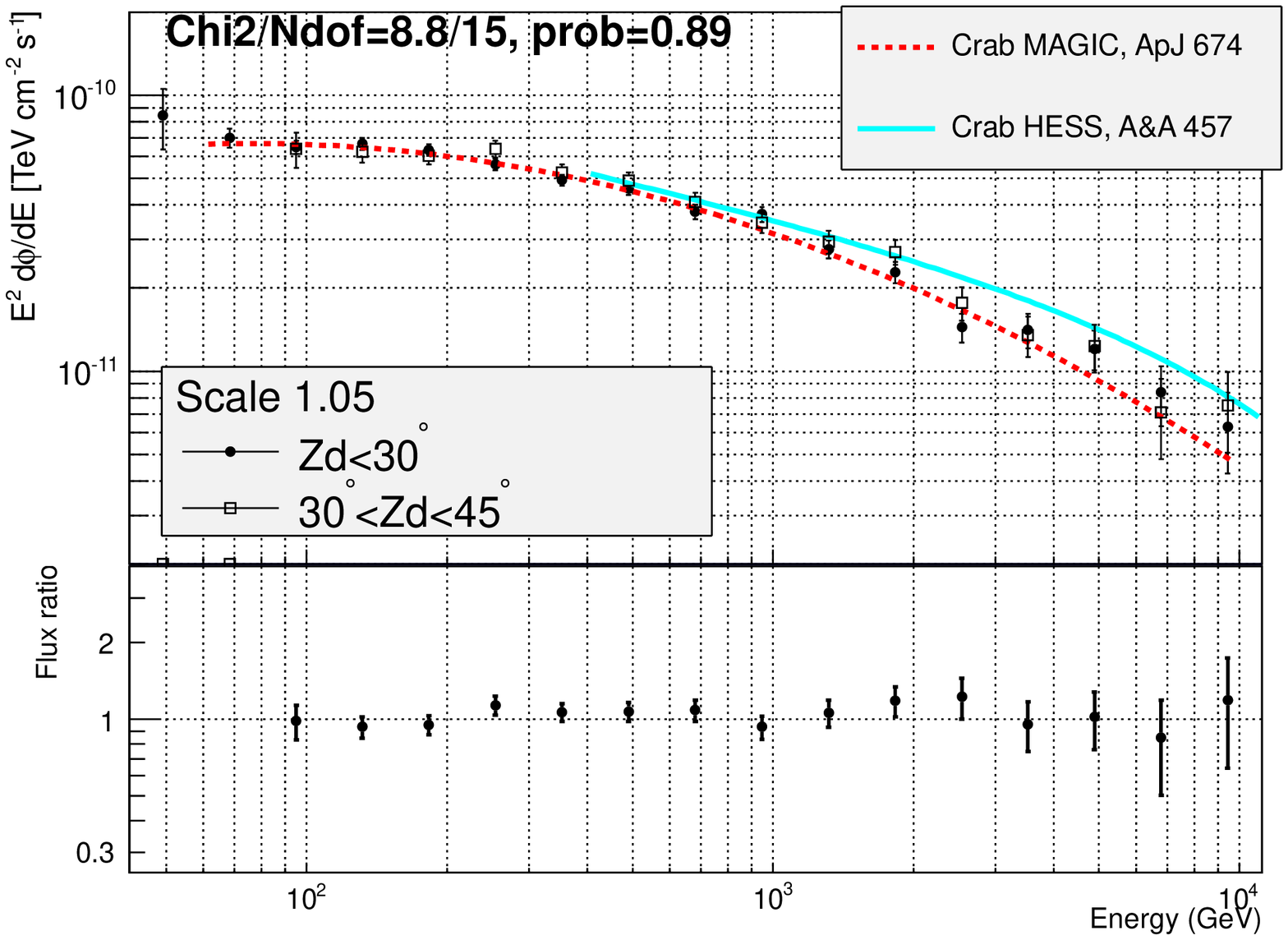}
\includegraphics[width=0.49\textwidth]{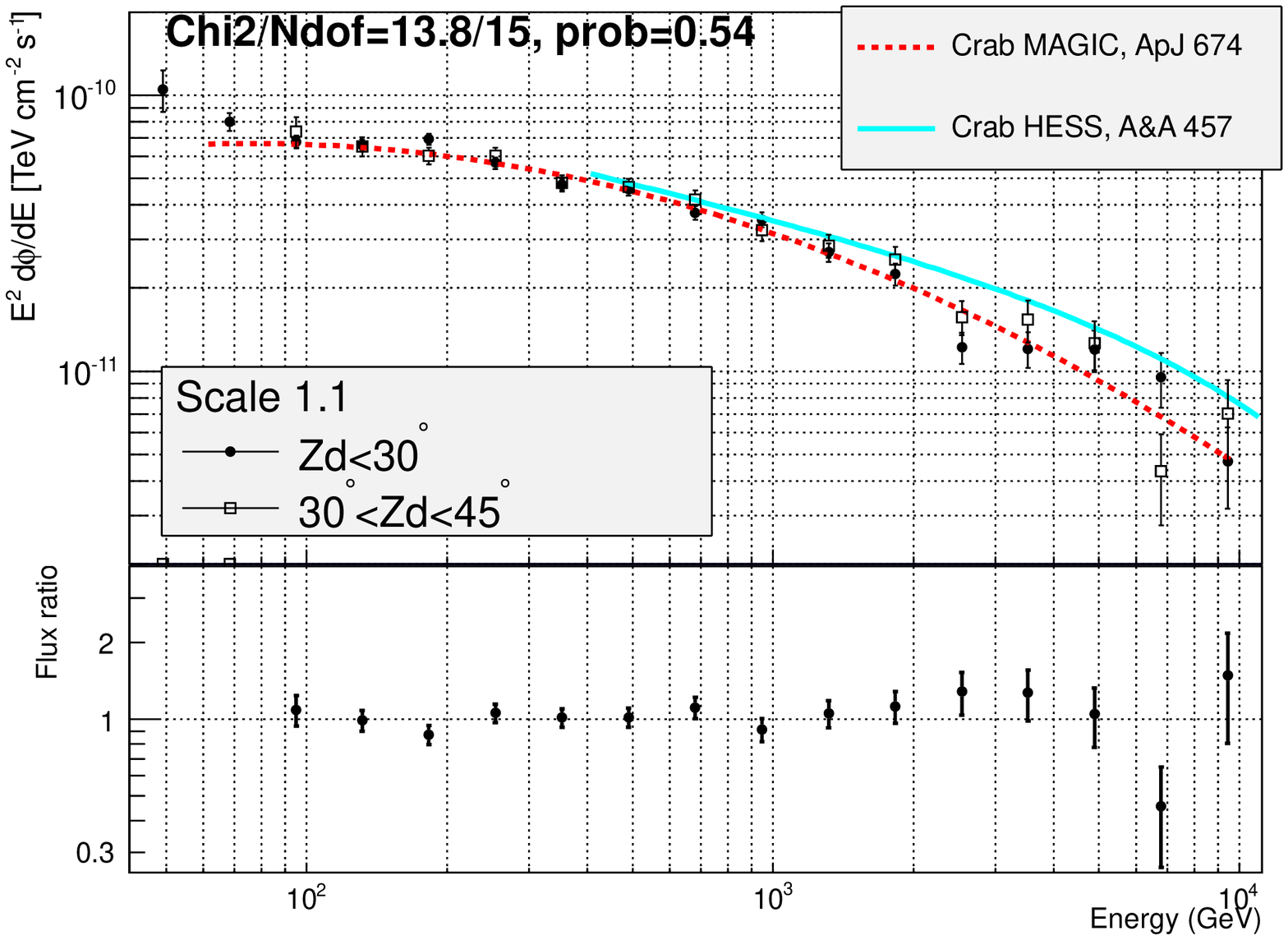}
\caption{
Spectra of the Crab Nebula obtained with the low (full circles) and medium (empty squares) zenith angle samples for MC with different energy scales (-25\%, -10\%, -5\%, no scale, +5\%, +10\%, see header of  legends).
The bottom panel shows the ratio of the medium zenith angle flux to the low zenith angle flux.
As a reference two historical Crab Nebula spectra (MAGIC, \cite{magic_crab}, and H.E.S.S. \cite{ah06}) are plotted with a dashed and a solid line. 
In the top part of individual panels we report the value of $\chi^2$, number of degrees of freedom and corresponding probability computed between the low and medium zenith angle spectra.
}\label{fig:syst_zeniths}
\end{figure*}

As the effect is most pronounced at the threshold we investigate the uncertainty of the light scale of the MAGIC telescopes by shifting the energy threshold in two different ways. 
We perform a set of full analyses using MC simulations with the light scale artificially shifted by -25\%, -10\%, -5\%, +5\%, +10\% and +25\%.
In the first scenario, we compare the reconstructed flux at a given energy for low and medium zenith angles (see Fig.~\ref{fig:syst_zeniths}). 
For the sake of comparison, we show in Fig.~\ref{fig:syst_zeniths} the two most spread apart historical Crab Nebula spectra observed by IACTs. 
For the low zenith angle spectrum, as the MC light scale is increased, an increasing pile-up effect shows up at low energies.
However, as observations at higher zenith angle do not reach such lowest energies, and at the energies of $\sim100\,$GeV, both spectra are consistent, such an overestimated light scale cannot be excluded aprori. 
On the other hand, when the MC light scale is decreased, there is a clear difference between the low and medium zenith angle samples at low energies. 
Moreover, even while the $\chi^2$ value computed between the two spectra give acceptable probabilities of $\sim 16\%$ and $\sim 10\%$ for down scaling by 10\% and 25\% respectively, there is a clear structure in the flux ratio plot. 

\begin{figure*}[pt]
%\centering 
\includegraphics[width=0.49\textwidth]{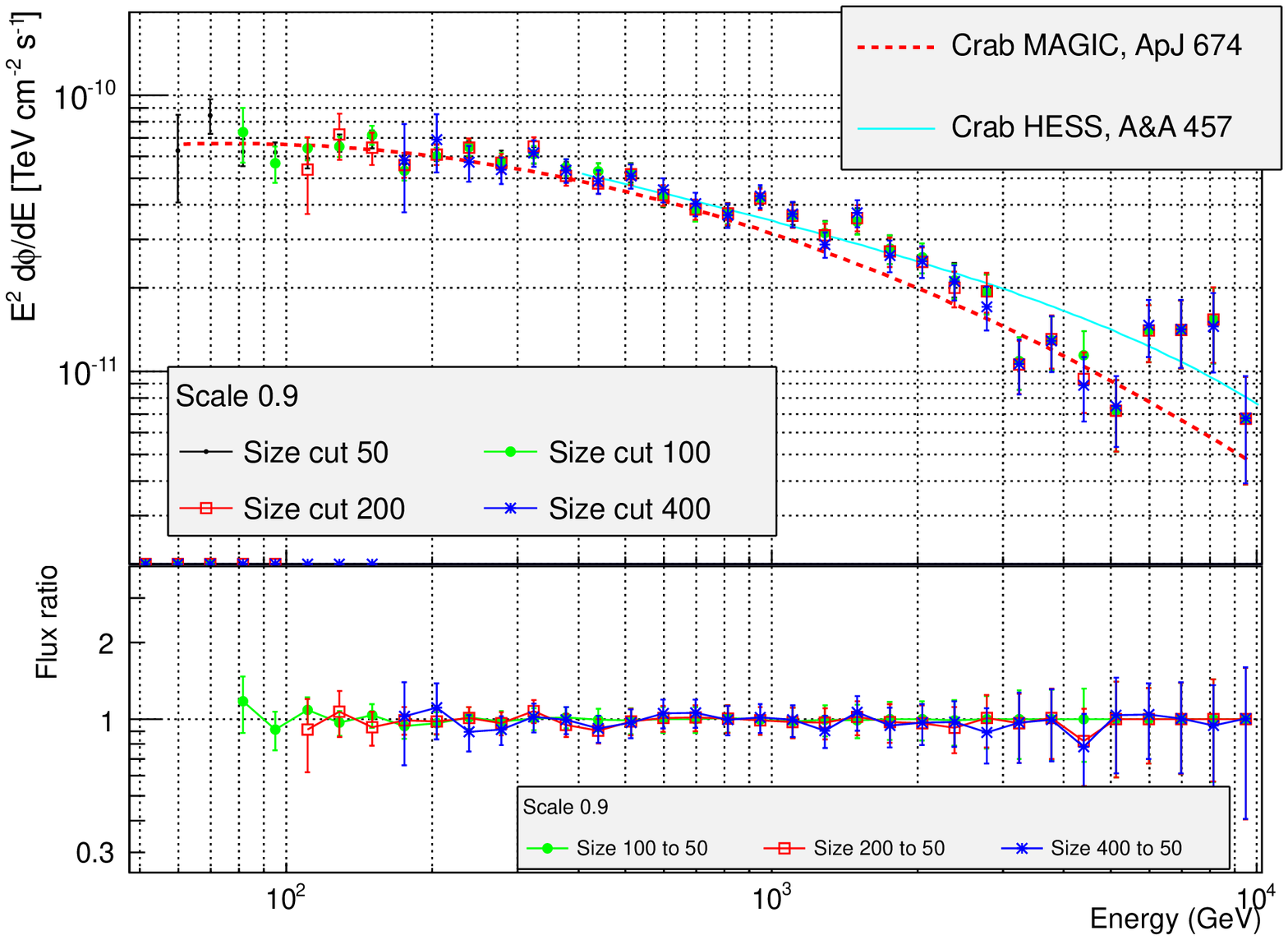}
\includegraphics[width=0.49\textwidth]{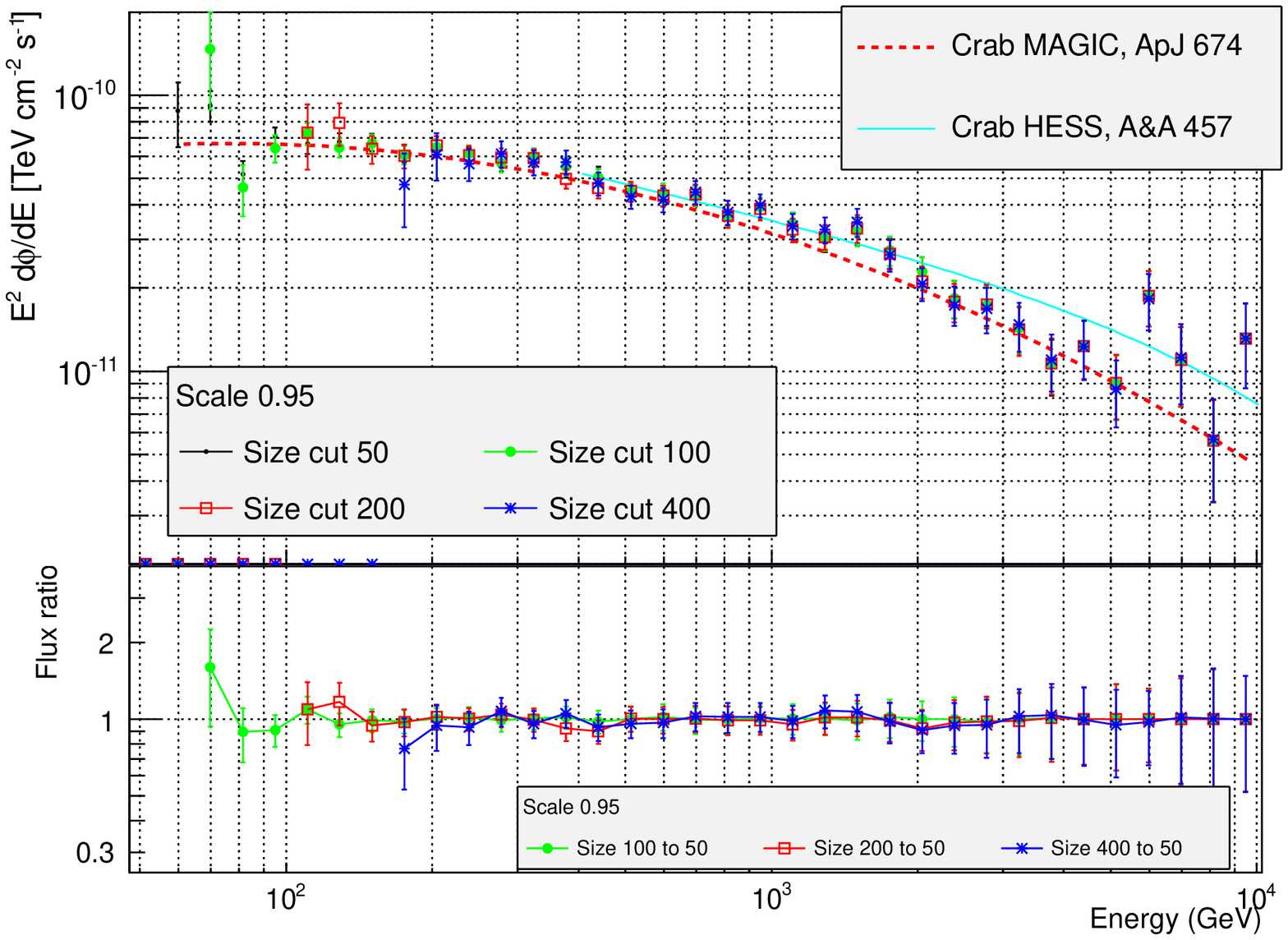}\\
\includegraphics[width=0.49\textwidth]{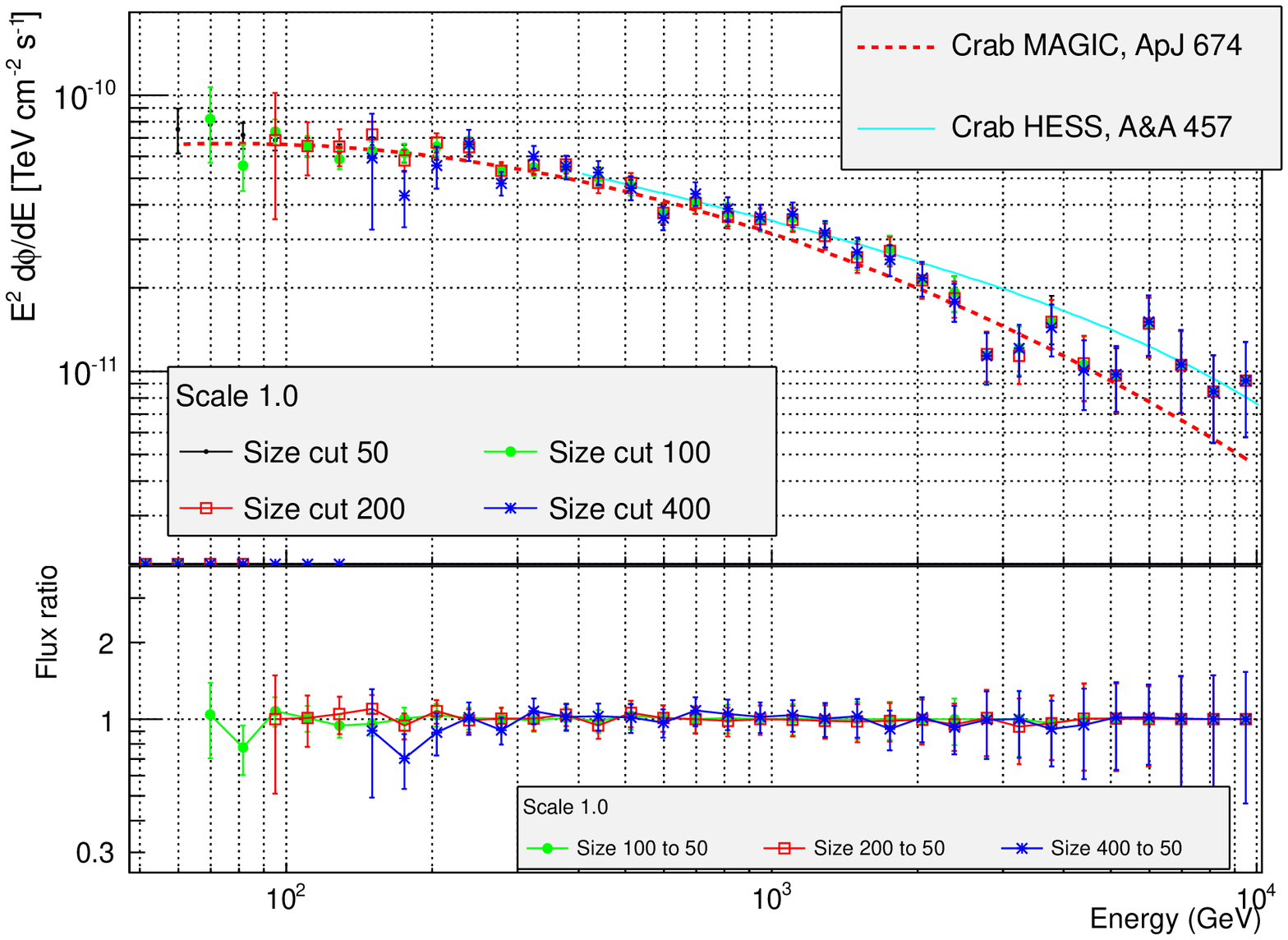}
\includegraphics[width=0.49\textwidth]{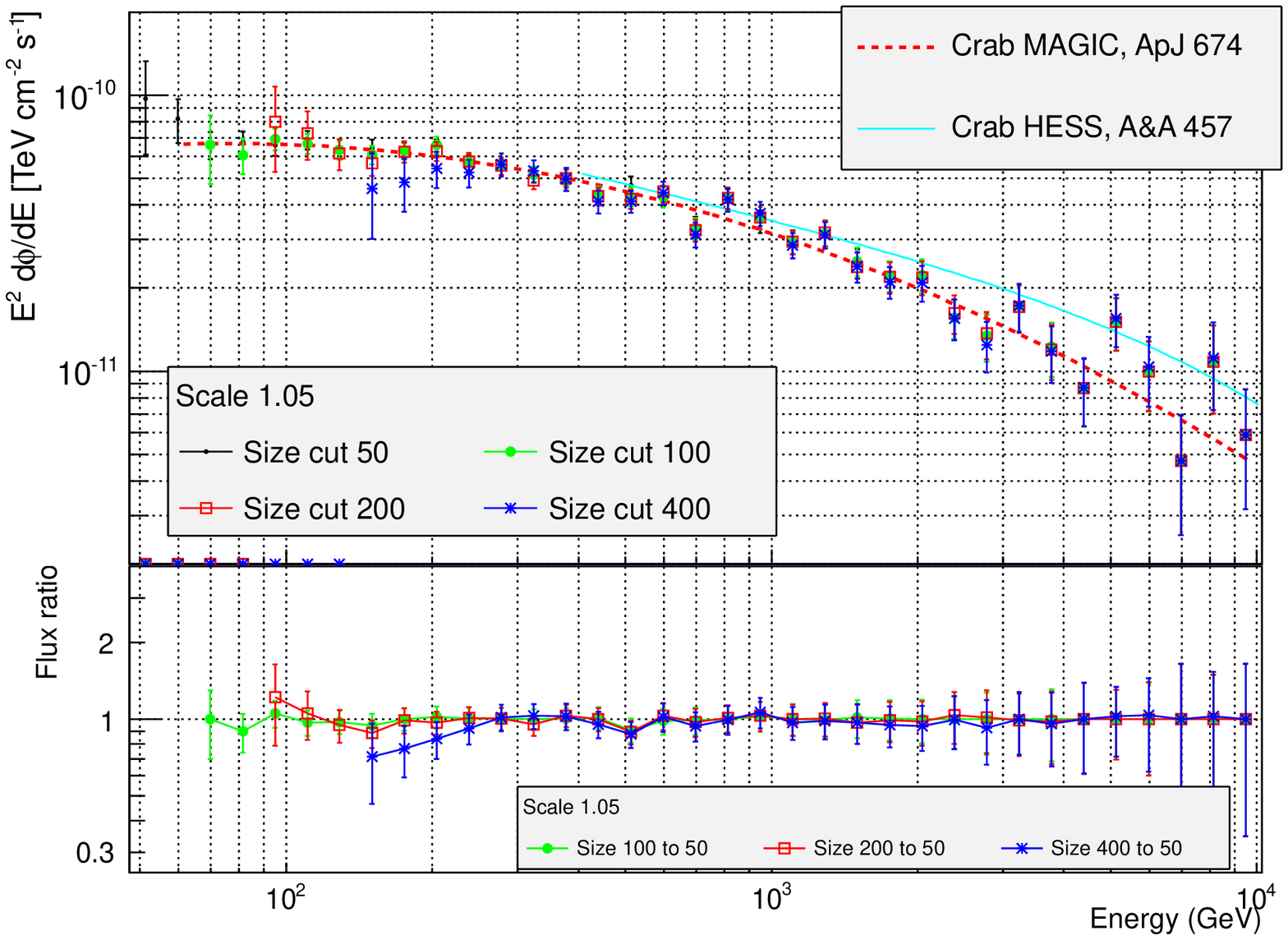}\\
\includegraphics[width=0.49\textwidth]{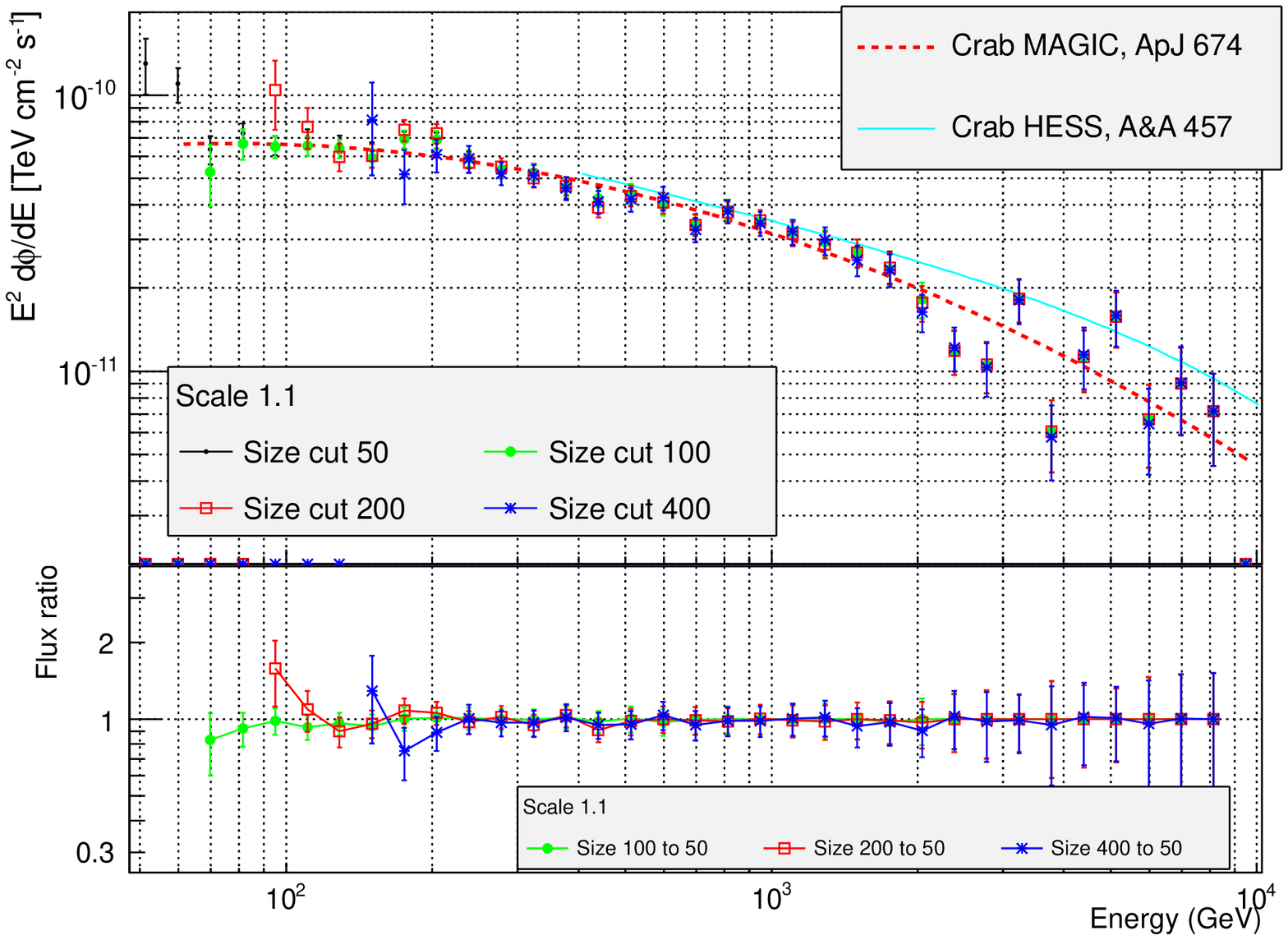}
\includegraphics[width=0.49\textwidth]{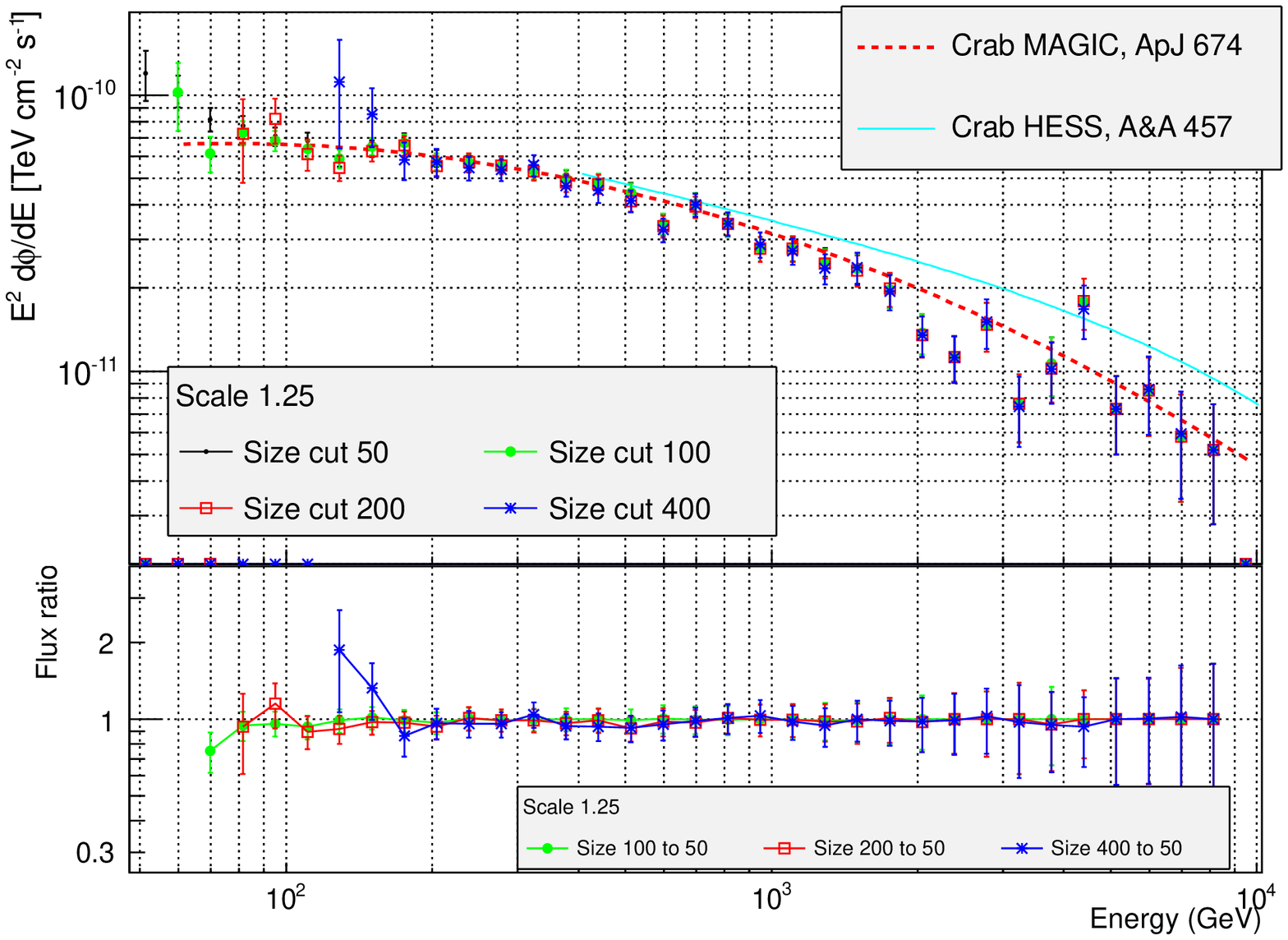}
\caption{
Spectra of the Crab Nebula obtained with the low zenith angle sample for different $Size$ cuts and with scaled MC:  -10\%, -5\%, no scale, +5\%, +10\% and +25\% (see the legend header in different panels). 
The spectra are obtained with $Size$ cuts of 50\,phe (black dots), 100\,phe (green full circles), 200\,phe (red empty squares) and 400\,phe (blue asterisks).
At the bottom, smaller sub-panels with the ratio between a spectrum with a given $Size$ cut and the reference one ($>50\,$phe) is shown.
As a reference, two historical Crab Nebula spectra (MAGIC, \cite{magic_crab}, and H.E.S.S. \cite{ah06}) are plotted with a dashed and solid line. 
For better visibility the data points are joined with broken dotted lines.
}\label{fig:syst_sizecuts}
\end{figure*}

A second way to artificially increase the threshold of the MAGIC telescopes is to use a higher \emph{Size} cut.
In Fig.~\ref{fig:syst_sizecuts} we show the Crab Nebula spectra for a low zenith angle sample obtained for \emph{Size} greater than 50, 100, 200 and 400\,phe. 
As before, we perform a full analysis of the Crab Nebula data for each of the MC samples with artificially shifted light scale. 
We compute the ratio of the flux obtained with a given cut in the $Size$ parameter to the reference cut of 50\,phe.
Note however that the plotted statistical uncertainties of the flux ratio are overestimated as the points between the spectra with different cuts in $Size$ are strongly correlated (the same Crab Nebula sample was used to obtain each of them).
In this case, the situation is opposite to the one with the energy threshold varying with the zenith angle of the observations.
For the underestimated light scale the spectra at different size cuts are still consistent between each other, thus no constraint on the absolute light scale can be drawn.
On the other hand, for the overestimated light scale, the flux ratio while going to lower energies has a characteristic V-shape (best visible with red curves in Fig.~\ref{fig:syst_sizecuts}). 
Going to lower energies it first slightly drops (due to the direct effect on the collection area) and then sharply increases (consistent with a pile-up effect from the bias in energy estimation below the energy threshold). 
Combining both methods we validate that the systematic uncertainty of the absolute energy scale of the MAGIC telescopes is below 15\%, similarly to the one expected in \cite{magic_stereo}.

\subsection{Systematic uncertainty in flux normalization and slope}
Due to better camera and trigger homogeneity after the upgrade and also lack of dead time in the readout the total systematic uncertainity in the flux normalization is slightly lower than evaluated in \cite{magic_stereo}.
We checked the contribution of the mismatch in cut efficiencies to the systematic uncertainty and estimate it to be $<12\%$ in the whole investigated energy range, compatible with the 10-15\% uncertainties due to analysis and data/MC discrepancies reported in \cite{magic_stereo}. 
The effect on the spectral indices was smaller than the statistical uncertainty obtained with this data sample.
We estimate the uncertainty on the flux normalization to be 18\% at low energies ($\lesssim100\,$GeV) and 11\% in the energy range of a few hundred GeV.
At the highest energies, $\gtrsim 1\,$TeV, due to more pronounced MC/data mismatches the systematic uncertainty is a bit higher, namely 16\%.
The systematic uncertainity on the reconstructed spectral slope of the sources is still $\pm0.15$ for a source with signal to background ratio of at least 25\%. 

\subsection{Night to night systematic uncertainty} 
The total systematic uncertainty estimated in the previous section is a proper quantity to be used while comparing the MAGIC observations with the data of other instruments or with theoretical predictions. 
However, a significant fraction of the systematic uncertainty is nearly constant and will affect all the MAGIC data in the same way. 
On the other hand, part of the systematic uncertainity (e.g. atmospheric transmission at a given day or small changes in the optical PSF) will vary from one night to another resulting in slightly different estimations of the flux even from a steady source.  
In order to estimate this remaining, relative systematic uncertainty we use the method of \cite{magic_stereo}.
We divided our data into sub-samples and compute the flux for each of them.
Then we compare the standard deviation, $\sigma_{F}$, of the distribution of the reconstructed fluxes with the typical uncertainty of individual points, $\delta F$, to determine the ``excess RMS''.
\begin{figure}[tb!]
\centering 
\includegraphics[width=0.49\textwidth]{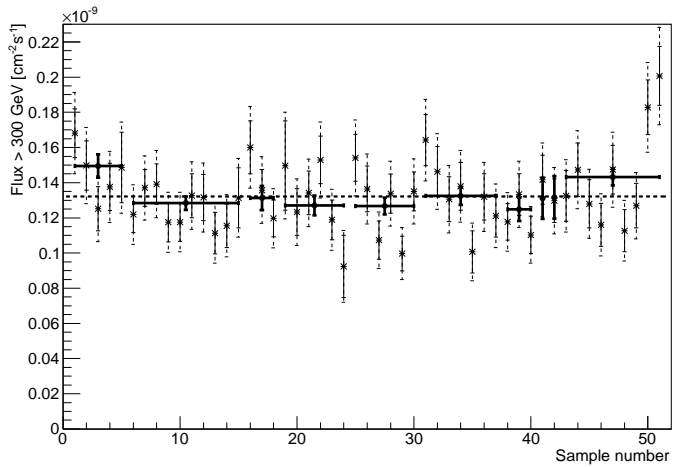}
\caption{
Integrated fluxes above 300$\,$GeV of the Crab Nebula for different data runs ($\sim 20$\,min) as a function of the data sample number are shown with thin lines and asterisks. 
The thick lines show the corresponding night by night fluxes (vertical error bars show the data span). 
The second set of dashed error bars for the thin lines shows the statistical uncertainty  added in quadrature with 11\% systematic uncertainty.
The dashed vertical line show the mean flux obtained from the total data sample.  
}\label{fig:lc}
\end{figure}

We first divide the data according to data runs. 
Each data run is normally 20-min long, and correspond to one wobble pointing. 
With such time binning the integrated flux above 300$\,$GeV is computed with a statistical precision of about 10\% (see Fig.\ref{fig:lc}).
Three short data runs (with a much higher relative statistical uncertainty of the flux estimation, $>20\%$) were removed from this study. 
Fitting the points with a constant we obtain $\chi^2/N_{dof}=106/50$, corresponding to a fit probability of $6.7\cdot 10^{-6}$.
The spread of the individual points is $\sigma_{F}\sim 15\%$, suggesting a relative run-to-run systematic uncertainty of $\sim 11\%$.
Note, that while so far there was no variability observed from the Crab Nebula, it is not excluded that a small intrinsic variability, below the accuracy of the current IACTs is present at the source itself.
Therefore, the values derived with this method can be treated as conservative estimates of the run to run systematic uncertainty of the MAGIC telescopes.
Adding such a systematic uncertainty of $11\%$ in quadrature to the statistical uncertainties of the individual points (see the dashed error bars in Fig.\ref{fig:lc}), and performing a fit with a constant, we obtain that the $\chi^2/N_{dof}=46/50$, nominally corresponding to probability of 63\%. 

In the case of night-by-night binning $\delta F$ is of the order of 5\%.
Interestingly, this light curve can be fitted with a constant flux with $\chi^2/N_{dof}=16.5/9$, which corresponds to a barely acceptable probability of 5.7\%.
We conservatively conclude that the relative systematic uncertainties are of the order of 11\%.
This value is similar to the values before the upgrade of the MAGIC telescopes  \cite{magic_stereo} as well as to the one obtained by the H.E.S.S telescopes \citep{ah06}.
It is plausible that most of this uncertainty is due to the atmospheric variations.

%% %-----------------------------------------------------------------------------

\section{Conclusions}\label{sec:concl}
The upgrade of the readout and one of the cameras of the MAGIC telescopes have significantly improved their performance. 
The trigger threshold for low zenith angle observations is $\sim 50\,$GeV. 
With the 15h sample of Crab Nebula data, its spectrum could be reconstructed between 65\,GeV and $13.5$\,TeV.
Within systematic uncertainties it is consistent with previous measurements of the Crab Nebula with IACTs. 
The best performance of the MAGIC telescopes is achieved at medium energies, at a few hundred GeV.
At those energies the images are sufficiently large to provide enough information for efficient reconstruction, while the rapidly falling power-law spectrum of the Crab Nebula still provides enough statistics.
The energy resolution at these medium energies is as good as 16\% with a negligible bias and the angular resolution is $\lesssim 0.07^\circ$. 
The sensitivity above 220\,GeV is $(0.66\pm0.03)\%$ of C.U. for 50h of observations. 
At the lowest energies, below $100\,$GeV, the performance has improved drastically, reducing the needed observation time by a factor of 2.5.
The larger trigger region and improved pixelization of the MAGIC-I camera have improved also the off-axis performance.
A source with a Crab Nebula like spectrum, but 80 times weaker, can be detected at the offset of $1^\circ$ within 50\,h of observations, making the MAGIC telescopes capable of efficient sky scans. 

The performance of the MAGIC telescopes at medium zenith angles, $30-45^\circ$, is mostly similar to the one at low zenith angles. 
The higher threshold, however, significantly degraded all the performance parameters below $\sim200$\,GeV.
For the highest energies, above a few TeV, better performance is achieved with observations at medium zenith angles.

We revised different sources of systematic uncertainities after the upgrade and studied in detail the uncertainty in the energy threshold. 
The larger trigger region has allowed us to lower the systematics connected with the background estimation by a factor of 2. 
From comparisons of reconstructed SEDs of the Crab Nebula for different energy thresholds we validated the systematic uncertainty in the energy scale to be below $15\%$.
The systematic uncertainty on the flux normalization was estimated to be 11-18\%, and on the spectral slope $\pm0.15$. 
The part of the systematic uncertainty which can change from one observation run to another is estimated to be about 11\%.

Thanks to the improvement in the performance achieved after the upgrade, the MAGIC telescopes have reached an unprecedented sensitivity.
Since then, five new VHE $\gamma$-ray sources have already been discovered by MAGIC.
Among them, 3C 58, is the least luminous pulsar wind nebula so far detected in the VHE $\gamma$-rays \citep{magic_3c58}.

\appendix
\section{Tables}
In this appendix we report, for easy reference, the numerical values of energy resolution and bias, angular resolution and sensitivity, together with additional information, such as corresponding rates of $\gamma$-rays.

\begin{table}[t]
\centering
\begin{tabular}{c|c|c|c}
$E [GeV]$ & bias [\%] & $\sigma$ [\%] & RMS [\%] \\\hline\hline
47 - 75 & $24.6 \pm 0.9$ & $21.8 \pm 1.1$ & $22.5 \pm 0.4$\\\hline
75 - 119 & $7.1 \pm 0.4$ & $19.8 \pm 0.5$ & $20.9 \pm 0.2$\\\hline
119 - 189 & $-0.1 \pm 0.3$ & $18.0 \pm 0.3$ & $21.3 \pm 0.2$\\\hline
189 - 299 & $-1.5 \pm 0.3$ & $16.8 \pm 0.3$ & $20.49 \pm 0.18$\\\hline
299 - 475 & $-2.2 \pm 0.2$ & $15.5 \pm 0.2$ & $20.20 \pm 0.17$\\\hline
475 - 753 & $-2.1 \pm 0.2$ & $14.8 \pm 0.2$ & $20.12 \pm 0.18$\\\hline
753 - 1194 & $-1.4 \pm 0.2$ & $15.4 \pm 0.2$ & $21.3 \pm 0.2$\\\hline
1194 - 1892 & $-1.8 \pm 0.3$ & $16.1 \pm 0.3$ & $21.3 \pm 0.2$\\\hline
1892 - 2999 & $-2.3 \pm 0.4$ & $18.1 \pm 0.4$ & $23.2 \pm 0.3$\\\hline
2999 - 4754 & $-1.7 \pm 0.4$ & $19.6 \pm 0.5$ & $25.1 \pm 0.3$\\\hline
4754 - 7535 & $-2.6 \pm 0.6$ & $21.9 \pm 0.6$ & $26.5 \pm 0.4$\\\hline
7535 - 11943 & $-2.1 \pm 0.8$ & $22.7 \pm 0.9$ & $26.8 \pm 0.5$\\\hline
11943 - 18928 & $-6.7 \pm 0.8$ & $20.7 \pm 0.9$ & $24.4 \pm 0.5$\\\hline
\end{tabular}
\caption{
Energy resolution and bias obtained from a low zenith angle ($0-30^\circ$) MC sample.
The individual columns report: 
$E$ - energy range,
bias and $\sigma$ - mean and standard deviation of a Gaussian fit to $(E_{est}-E_{true})/E_{true}$ distribution, 
RMS - standard deviation obtained directly from this distribution.}
\label{tab:enreslz}
\end{table}

\begin{table}[t]
\centering
\begin{tabular}{c|c|c|c}
$E [GeV]$ & bias [\%] & $\sigma$ [\%] & RMS [\%] \\\hline\hline
47 - 75 & $45.8 \pm 1.8$ & $23 \pm 2$ & $26.6 \pm 1.0$\\\hline
75 - 119 & $18.9 \pm 0.7$ & $21.3 \pm 0.8$ & $23.7 \pm 0.4$\\\hline
119 - 189 & $3.8 \pm 0.4$ & $18.9 \pm 0.4$ & $22.2 \pm 0.3$\\\hline
189 - 299 & $-2.0 \pm 0.3$ & $18.2 \pm 0.3$ & $23.2 \pm 0.2$\\\hline
299 - 475 & $-3.8 \pm 0.3$ & $17.5 \pm 0.3$ & $22.3 \pm 0.2$\\\hline
475 - 753 & $-2.6 \pm 0.3$ & $16.8 \pm 0.3$ & $23.0 \pm 0.2$\\\hline
753 - 1194 & $-1.5 \pm 0.3$ & $16.6 \pm 0.3$ & $23.2 \pm 0.2$\\\hline
1194 - 1892 & $-0.4 \pm 0.3$ & $16.6 \pm 0.3$ & $24.5 \pm 0.3$\\\hline
1892 - 2999 & $-0.2 \pm 0.3$ & $17.3 \pm 0.3$ & $25.2 \pm 0.3$\\\hline
2999 - 4754 & $0.5 \pm 0.4$ & $18.8 \pm 0.4$ & $26.0 \pm 0.3$\\\hline
4754 - 7535 & $0.9 \pm 0.5$ & $20.5 \pm 0.5$ & $28.0 \pm 0.4$\\\hline
7535 - 11943 & $-0.1 \pm 0.7$ & $22.3 \pm 0.7$ & $29.2 \pm 0.5$\\\hline
11943 - 18928 & $0.7 \pm 0.8$ & $22.7 \pm 0.8$ & $29.5 \pm 0.6$\\\hline
\end{tabular}
\caption{
Energy resolution and bias obtained from a medium zenith angle ($30-45^\circ$) MC sample. Columns as in Table \ref{tab:enreslz}.}
\label{tab:enresmz}
\end{table}

\begin{table}[t]
\centering
\tabcolsep=0.11cm
\begin{tabular}{c|c|c|c|c}
$E$ &  \multicolumn{2}{c|}{Zenith angle $<30^\circ$} &  \multicolumn{2}{c}{Zenith angle $30-45^\circ$} \\
~[GeV] & $\Theta_{Gaus}[^\circ]$ & $\Theta_{0.68}[^\circ]$ & $\Theta_{Gaus}[^\circ]$ & $\Theta_{0.68}[^\circ]$  \\\hline\hline 
95 & $0.087\pm0.004$ & $0.157_{-0.007}^{+0.007}$ & $0.088\pm0.013$ & $0.129_{-0.021}^{+0.009}$ \rule{0pt}{10pt}\\[0.5ex]\hline
150 & $0.075\pm0.002$ & $0.135_{-0.005}^{+0.005}$ & $0.078\pm0.005$ & $0.148_{-0.013}^{+0.017}$ \rule{0pt}{10pt}\\[0.5ex]\hline
238 & $0.067\pm0.001$ & $0.108_{-0.003}^{+0.004}$ & $0.072\pm0.003$ & $0.120_{-0.007}^{+0.009}$ \rule{0pt}{10pt}\\[0.5ex]\hline
378 & $0.058\pm0.001$ & $0.095_{-0.003}^{+0.004}$ & $0.063\pm0.003$ & $0.097_{-0.006}^{+0.008}$ \rule{0pt}{10pt}\\[0.5ex]\hline
599 & $0.052\pm0.001$ & $0.081_{-0.003}^{+0.003}$ & $0.054\pm0.003$ & $0.083_{-0.007}^{+0.007}$ \rule{0pt}{10pt}\\[0.5ex]\hline
949 & $0.046\pm0.001$ & $0.073_{-0.003}^{+0.004}$ & $0.052\pm0.002$ & $0.082_{-0.005}^{+0.006}$ \rule{0pt}{10pt}\\[0.5ex]\hline
1504 & $0.044\pm0.001$ & $0.071_{-0.003}^{+0.005}$ & $0.046\pm0.002$ & $0.077_{-0.004}^{+0.007}$ \rule{0pt}{10pt}\\[0.5ex]\hline
2383 & $0.042\pm0.002$ & $0.067_{-0.005}^{+0.006}$ & $0.045\pm0.003$ & $0.068_{-0.006}^{+0.010}$ \rule{0pt}{10pt}\\[0.5ex]\hline
3777 & $0.042\pm0.003$ & $0.065_{-0.004}^{+0.011}$ & $0.039\pm0.004$ & $0.061_{-0.008}^{+0.011}$ \rule{0pt}{10pt}\\[0.5ex]\hline
5986 & $0.041\pm0.004$ & $0.062_{-0.011}^{+0.012}$ & $0.038\pm0.006$ & $0.059_{-0.011}^{+0.031}$ \rule{0pt}{10pt}\\[0.5ex]\hline
9487 & $0.040\pm0.005$ & $0.056_{-0.012}^{+0.062}$ & $0.046\pm0.009$ & $0.055_{-0.005}^{+0.209}$ \rule{0pt}{10pt}\\[0.5ex]\hline
\end{tabular}
\caption{
Angular resolution $\Theta_{Gaus}$ and $\Theta_{0.68}$ of the MAGIC telescopes after the upgrade as a function of the estimated energy $E$, obtained with the Crab Nebula data sample.
$\Theta_{Gaus}$ is computed as a sigma of a 2D Gaussian fit.
$\Theta_{0.68}$ is the 68\% containment radius of the $\gamma$-ray excess.
}\label{tab:angres}
\end{table}

\begin{table*}[t!]
\centering
\begin{tabular*}{0.99\textwidth}{@{\extracolsep{\fill}} c|c|c|c|c|c|c|c}
$E_{\rm thresh.}$ & $\gamma$-rate & bkg-rate & $S_{Nex/\!\sqrt{Nbkg}}$ & $S_{\rm Li\&Ma, 1Off}$ & $S_{\rm Li\&Ma, 3Off}$ & $S_{\rm Li\&Ma, 5Off}$ & $S_{Nex/\!\sqrt{Nbkg}}$  \\
~[GeV] & [min$^{-1}$] & [min$^{-1}$] & [\%C.U.] & [\%C.U.] & [\%C.U.] & [\%C.U.]&[$10^{-13}$$\rm cm^{-2} s^{-1}$] \\\hline\hline
84  & $19.1 \pm 0.2$  & $8.73 \pm 0.07$  & $2.29 \pm 0.03$  & $2.29 \pm 0.03$  & $2.29 \pm 0.03$  & $2.29 \pm 0.03$  & $156.5 \pm 1.7$ \\ \hline
86  & $18.8 \pm 0.2$  & $7.80 \pm 0.06$  & $2.07 \pm 0.02$  & $2.07 \pm 0.03$  & $2.07 \pm 0.03$  & $2.07 \pm 0.03$  & $137.1 \pm 1.5$ \\ \hline
104  & $16.88 \pm 0.19$  & $4.88 \pm 0.05$  & $1.445 \pm 0.015$ & $1.71 \pm 0.02$  & $1.45 \pm 0.02$  & $1.45 \pm 0.02$  & $75.9 \pm 0.8$ \\ \hline
146  & $6.17 \pm 0.10$  & $0.320 \pm 0.013$  & $0.84 \pm 0.02$  & $1.25 \pm 0.03$  & $1.00 \pm 0.02$  & $0.95 \pm 0.02$  & $28.6 \pm 0.8$ \\ \hline
218  & $3.63 \pm 0.07$  & $0.070 \pm 0.006$  & $0.66 \pm 0.03$  & $1.06 \pm 0.04$  & $0.83 \pm 0.03$  & $0.78 \pm 0.03$  & $13.4 \pm 0.7$ \\ \hline
289  & $2.94 \pm 0.07$  & $0.032 \pm 0.004$  & $0.56 \pm 0.04$  & $0.93 \pm 0.05$  & $0.72 \pm 0.04$  & $0.67 \pm 0.04$  & $7.6 \pm 0.5$ \\ \hline
404  & $3.05 \pm 0.07$  & $0.030 \pm 0.004$  & $0.51 \pm 0.04$  & $0.87 \pm 0.04$  & $0.67 \pm 0.04$  & $0.63 \pm 0.03$  & $4.4 \pm 0.3$ \\ \hline
523  & $2.51 \pm 0.06$  & $0.023 \pm 0.003$  & $0.55 \pm 0.04$  & $0.95 \pm 0.05$  & $0.72 \pm 0.04$  & $0.68 \pm 0.05$  & $3.2 \pm 0.3$ \\ \hline
803  & $1.59 \pm 0.05$  & $0.0109 \pm 0.0010$  & $0.60 \pm 0.03$  & $1.12 \pm 0.04$  & $0.84 \pm 0.03$  & $0.78 \pm 0.03$  & $1.78 \pm 0.10$ \\ \hline
1233  & $0.95 \pm 0.04$  & $0.0062 \pm 0.0007$  & $0.75 \pm 0.05$  & $1.53 \pm 0.06$  & $1.11 \pm 0.05$  & $1.02 \pm 0.05$  & $1.12 \pm 0.08$ \\ \hline
1935  & $0.55 \pm 0.03$  & $0.0053 \pm 0.0011$  & $1.20 \pm 0.15$  & $2.50 \pm 0.16$  & $1.80 \pm 0.13$  & $1.66 \pm 0.15$  & $0.83 \pm 0.10$ \\ \hline
2938  & $0.31 \pm 0.02$  & $0.0027 \pm 0.0008$  & $1.6 \pm 0.2$  & $3.7 \pm 0.3$  & $2.6 \pm 0.3$  & $2.3 \pm 0.2$  & $0.51 \pm 0.08$ \\ \hline
4431  & $0.160 \pm 0.016$  & $0.0022 \pm 0.0006$  & $2.7 \pm 0.5$  & $6.6 \pm 0.6$  & $4.5 \pm 0.5$  & $4.1 \pm 0.4$  & $0.41 \pm 0.07$ \\ \hline
6718  & $0.078 \pm 0.011$  & $0.0018 \pm 0.0010$  & $4.9 \pm 1.6$  & $12.9 \pm 1.7$  & $8.7 \pm 1.5$  & $7.8 \pm 1.4$  & $0.34 \pm 0.11$ \\ \hline
8760  & $0.046 \pm 0.008$  & $0.0005 \pm 0.0005$  & $7.2 \pm 1.3$  & $17 \pm 2$  & $10.4 \pm 1.7$  & $9.0 \pm 1.8$  & $0.30 \pm 0.05$ \\ \hline
\end{tabular*}
\caption{
Integral sensitivity of the MAGIC telescopes obtained with the low zenith angle Crab Nebula data sample above a given energy threshold, $E_{\rm thresh.}$.
The sensitivity is calculated as $N_{\rm excess}/\sqrt{N_{\rm bkg}}=5$  ($S_{Nex/\!\sqrt{Nbkg}}$), or according to the 5$\sigma$ significance obtained from \citet{lm83} (using 1, 3 or 5 background regions,  $S_{\rm Li\&Ma, 1Off}$, $S_{\rm Li\&Ma, 3Off}$ and $S_{\rm Li\&Ma, 5Off}$). 
The sensitivity is computed for 50$\,$h of observation time with the additional conditions $N_{\rm excess}>10$, $N_{\rm excess}> 0.05 N_{\rm bkg}$ ($S_{Nex/\!\sqrt{Nbkg},\, {\rm sys}}$).
The $\gamma$-rate and bkg-rate columns show the rate of $\gamma$ events from Crab Nebula and residual background respectively above the energy threshold of the sensitivity point. 
}\label{tab:intsens}
\end{table*}

\begin{table*}[t!]
\centering
\begin{tabular*}{0.99\textwidth}{@{\extracolsep{\fill}} c|c|c|c|c|c|c|c}
$E_{\rm thresh.}$ & $\gamma$-rate & bkg-rate & $S_{Nex/\!\sqrt{Nbkg}}$ & $S_{\rm Li\&Ma, 1Off}$ & $S_{\rm Li\&Ma, 3Off}$ & $S_{\rm Li\&Ma, 5Off}$ & $S_{Nex/\!\sqrt{Nbkg}}$  \\
~[GeV] & [min$^{-1}$] & [min$^{-1}$] & [\%C.U.] & [\%C.U.] & [\%C.U.] & [\%C.U.]&[$10^{-13}$ $\rm cm^{-2} s^{-1}$] \\\hline\hline
114  & $19.7 \pm 0.4$  & $7.68 \pm 0.10$  & $1.95 \pm 0.03$  & $1.95 \pm 0.04$  & $1.95 \pm 0.04$  & $1.95 \pm 0.04$  & $91.4 \pm 1.6$ \\ \hline
119  & $19.3 \pm 0.3$  & $6.83 \pm 0.10$  & $1.76 \pm 0.03$  & $1.77 \pm 0.03$  & $1.76 \pm 0.04$  & $1.76 \pm 0.04$  & $78.4 \pm 1.3$ \\ \hline
141  & $17.4 \pm 0.3$  & $4.23 \pm 0.08$  & $1.22 \pm 0.02$  & $1.55 \pm 0.03$  & $1.26 \pm 0.03$  & $1.22 \pm 0.03$  & $43.5 \pm 0.8$ \\ \hline
210  & $5.60 \pm 0.16$  & $0.216 \pm 0.017$  & $0.76 \pm 0.04$  & $1.15 \pm 0.05$  & $0.92 \pm 0.04$  & $0.86 \pm 0.04$  & $16.1 \pm 0.8$ \\ \hline
310  & $3.59 \pm 0.12$  & $0.070 \pm 0.010$  & $0.67 \pm 0.05$  & $1.07 \pm 0.07$  & $0.84 \pm 0.06$  & $0.79 \pm 0.06$  & $8.3 \pm 0.7$ \\ \hline
401  & $3.19 \pm 0.12$  & $0.051 \pm 0.008$  & $0.65 \pm 0.06$  & $1.05 \pm 0.08$  & $0.82 \pm 0.07$  & $0.77 \pm 0.06$  & $5.6 \pm 0.5$ \\ \hline
435  & $3.31 \pm 0.12$  & $0.053 \pm 0.009$  & $0.63 \pm 0.06$  & $1.03 \pm 0.08$  & $0.80 \pm 0.07$  & $0.75 \pm 0.06$  & $4.8 \pm 0.4$ \\ \hline
546  & $2.98 \pm 0.11$  & $0.042 \pm 0.008$  & $0.63 \pm 0.06$  & $1.03 \pm 0.08$  & $0.80 \pm 0.07$  & $0.75 \pm 0.06$  & $3.4 \pm 0.3$ \\ \hline
821  & $2.24 \pm 0.10$  & $0.023 \pm 0.002$  & $0.61 \pm 0.04$  & $1.06 \pm 0.05$  & $0.81 \pm 0.05$  & $0.76 \pm 0.04$  & $1.77 \pm 0.12$ \\ \hline
1262  & $1.19 \pm 0.07$  & $0.0086 \pm 0.0014$  & $0.71 \pm 0.07$  & $1.38 \pm 0.09$  & $1.02 \pm 0.08$  & $0.94 \pm 0.07$  & $1.02 \pm 0.10$ \\ \hline
1955  & $0.84 \pm 0.06$  & $0.0072 \pm 0.0013$  & $0.93 \pm 0.11$  & $1.84 \pm 0.13$  & $1.35 \pm 0.11$  & $1.24 \pm 0.10$  & $0.63 \pm 0.07$ \\ \hline
2891  & $0.60 \pm 0.05$  & $0.013 \pm 0.003$  & $1.7 \pm 0.2$  & $3.2 \pm 0.3$  & $2.4 \pm 0.2$  & $2.2 \pm 0.3$  & $0.58 \pm 0.08$ \\ \hline
4479  & $0.31 \pm 0.04$  & $0.0052 \pm 0.0017$  & $2.1 \pm 0.4$  & $4.5 \pm 0.6$  & $3.2 \pm 0.5$  & $3.0 \pm 0.4$  & $0.32 \pm 0.07$ \\ \hline
7133  & $0.14 \pm 0.02$  & $0.0017 \pm 0.0010$  & $2.7 \pm 0.9$  & $7.1 \pm 1.0$  & $4.8 \pm 0.9$  & $4.3 \pm 0.9$  & $0.17 \pm 0.06$ \\ \hline
\end{tabular*}
\caption{
Integral sensitivity of the MAGIC telescopes obtained with the medium zenith angle ($30-45^\circ$) Crab Nebula data sample above a given energy threshold $E_{\rm thresh.}$.
Columns as in Table~\ref{tab:intsens}.
}\label{tab:intsens2}
\end{table*}

\begin{table*}[t!]
\centering
\begin{tabular*}{0.99\textwidth}{@{\extracolsep{\fill}} c|c|c|c|c|c|c|c|c}
$E_{\min}$ & $E_{\max}$ & $\gamma$-rate & bkg-rate & $S_{Nex/\!\sqrt{Nbkg}}$ & $S_{\rm Li\&Ma, 1Off}$ & $S_{\rm Li\&Ma, 3Off}$ & $S_{\rm Li\&Ma, 5Off}$ & $S_{Nex/\!\sqrt{Nbkg}}$  \\
~[GeV] & ~[GeV] & [min$^{-1}$] & [min$^{-1}$] & [\%C.U.] & [\%C.U.] & [\%C.U.] & [\%C.U.]&\footnotesize{[$10^{-12}$ $\rm cm^{-2} s^{-1} TeV^{-1}$]} \\\hline\hline
63 & 100  & $3.01 \pm 0.13$  & $4.06 \pm 0.08$  & $6.7 \pm 0.2$  & $8.8 \pm 0.4$  & $7.1 \pm 0.3$  & $6.8 \pm 0.3$  & $730 \pm 30$ \\ \hline
100 & 158  & $4.29 \pm 0.12$  & $2.41 \pm 0.06$  & $3.31 \pm 0.12$  & $4.77 \pm 0.14$  & $3.87 \pm 0.11$  & $3.67 \pm 0.10$  & $137 \pm 5$ \\ \hline
158 & 251  & $3.37 \pm 0.08$  & $0.54 \pm 0.03$  & $2.00 \pm 0.08$  & $2.95 \pm 0.10$  & $2.38 \pm 0.08$  & $2.25 \pm 0.08$  & $30.5 \pm 1.3$ \\ \hline
251 & 398  & $1.36 \pm 0.05$  & $0.066 \pm 0.010$  & $1.72 \pm 0.15$  & $2.8 \pm 0.2$  & $2.16 \pm 0.16$  & $2.03 \pm 0.15$  & $9.3 \pm 0.8$ \\ \hline
398 & 631  & $1.22 \pm 0.04$  & $0.027 \pm 0.006$  & $1.23 \pm 0.16$  & $2.10 \pm 0.18$  & $1.61 \pm 0.18$  & $1.51 \pm 0.15$  & $2.3 \pm 0.3$ \\ \hline
631 & 1000  & $0.88 \pm 0.04$  & $0.0133 \pm 0.0018$  & $1.19 \pm 0.10$  & $2.18 \pm 0.12$  & $1.64 \pm 0.09$  & $1.53 \pm 0.11$  & $0.72 \pm 0.06$ \\ \hline
1000 & 1585  & $0.58 \pm 0.03$  & $0.0059 \pm 0.0007$  & $1.21 \pm 0.10$  & $2.48 \pm 0.11$  & $1.80 \pm 0.09$  & $1.66 \pm 0.09$  & $0.230 \pm 0.018$ \\ \hline
1585 & 2512  & $0.30 \pm 0.02$  & $0.0027 \pm 0.0005$  & $1.58 \pm 0.18$  & $3.8 \pm 0.2$  & $2.60 \pm 0.19$  & $2.36 \pm 0.18$  & $0.090 \pm 0.010$ \\ \hline
2512 & 3981  & $0.166 \pm 0.016$  & $0.0020 \pm 0.0005$  & $2.5 \pm 0.4$  & $6.2 \pm 0.5$  & $4.3 \pm 0.4$  & $3.8 \pm 0.4$  & $0.041 \pm 0.007$ \\ \hline
3981 & 6310  & $0.093 \pm 0.012$  & $0.0014 \pm 0.0003$  & $3.7 \pm 0.7$  & $10.2 \pm 1.0$  & $6.8 \pm 0.7$  & $6.1 \pm 0.7$  & $0.017 \pm 0.003$ \\ \hline
6310 & 10000  & $0.060 \pm 0.010$  & $0.0046 \pm 0.0015$  & $10 \pm 3$  & $22 \pm 3$  & $16 \pm 3$  & $15 \pm 2$  & $0.013 \pm 0.003$ \\ \hline
\end{tabular*}
\caption{
Differential sensitivity of the MAGIC telescopes obtained with the low zenith angle observations of Crab Nebula data sample.
The definitions of the sensitivities are as in Table~\ref{tab:intsens}.
The $\gamma$-rate and bkg-rate columns show the rate of $\gamma$ events from Crab Nebula and residual background respectively in the differential estimated energy bins. 
}\label{tab:diffsens}
\end{table*}
\begin{table*}[t!]
\centering
\begin{tabular*}{0.99\textwidth}{@{\extracolsep{\fill}} c|c|c|c|c|c|c|c|c}
$E_{\min}$ & $E_{\max}$ & $\gamma$-rate & bkg-rate & $S_{Nex/\!\sqrt{Nbkg}}$ & $S_{\rm Li\&Ma, 1Off}$ & $S_{\rm Li\&Ma, 3Off}$ & $S_{\rm Li\&Ma, 5Off}$ & $S_{Nex/\!\sqrt{Nbkg}}$  \\
~[GeV] & ~[GeV] & [min$^{-1}$] & [min$^{-1}$] & [\%C.U.] & [\%C.U.] & [\%C.U.] & [\%C.U.]&\footnotesize{[$10^{-12}$ $\rm cm^{-2} s^{-1} TeV^{-1}$]} \\\hline\hline
63 & 100  & $0.40 \pm 0.12$  & $2.92 \pm 0.11$  & $39 \pm 16$  & $56 \pm 16$  & $45 \pm 12$  & $43 \pm 11$  & $4200 \pm 1700$ \\ \hline
100 & 158  & $3.18 \pm 0.16$  & $2.89 \pm 0.05$  & $4.9 \pm 0.4$  & $7.0 \pm 0.4$  & $5.7 \pm 0.3$  & $5.4 \pm 0.3$  & $202 \pm 15$ \\ \hline
158 & 251  & $2.67 \pm 0.19$  & $0.54 \pm 0.04$  & $2.52 \pm 0.19$  & $3.7 \pm 0.3$  & $3.0 \pm 0.3$  & $2.8 \pm 0.2$  & $38 \pm 3$ \\ \hline
251 & 398  & $2.86 \pm 0.13$  & $0.305 \pm 0.019$  & $1.76 \pm 0.14$  & $2.64 \pm 0.14$  & $2.11 \pm 0.11$  & $2.00 \pm 0.10$  & $9.5 \pm 0.8$ \\ \hline
398 & 631  & $1.76 \pm 0.12$  & $0.088 \pm 0.006$  & $1.5 \pm 0.2$  & $2.41 \pm 0.16$  & $1.90 \pm 0.14$  & $1.79 \pm 0.13$  & $2.8 \pm 0.4$ \\ \hline
631 & 1000  & $1.44 \pm 0.09$  & $0.038 \pm 0.002$  & $1.23 \pm 0.13$  & $2.04 \pm 0.12$  & $1.58 \pm 0.09$  & $1.48 \pm 0.10$  & $0.74 \pm 0.08$ \\ \hline
1000 & 1585  & $0.94 \pm 0.08$  & $0.0197 \pm 0.0016$  & $1.36 \pm 0.12$  & $2.38 \pm 0.16$  & $1.81 \pm 0.13$  & $1.69 \pm 0.13$  & $0.26 \pm 0.02$ \\ \hline
1585 & 2512  & $0.67 \pm 0.06$  & $0.0111 \pm 0.0015$  & $1.43 \pm 0.16$  & $2.7 \pm 0.2$ & $2.00 \pm 0.19$  & $1.85 \pm 0.18$  & $0.082 \pm 0.009$ \\ \hline
2512 & 3981  & $0.32 \pm 0.05$  & $0.0093 \pm 0.0012$  & $2.8 \pm 0.4$  & $5.3 \pm 0.7$  & $3.9 \pm 0.6$  & $3.7 \pm 0.5$  & $0.046 \pm 0.007$ \\ \hline
3981 & 6310  & $0.20 \pm 0.04$  & $0.0042 \pm 0.0017$  & $2.9 \pm 0.6$  & $6.4 \pm 1.2$  & $4.6 \pm 0.9$  & $4.2 \pm 0.9$  & $0.014 \pm 0.003$ \\ \hline
6310 & 10000 & $0.10 \pm 0.03$  & $0.0052 \pm 0.0002$  & $6.7 \pm 1.9$  & $14 \pm 3$  & $10 \pm 3$  & $9 \pm 2$  & $0.008 \pm 0.002$ \\ \hline
\end{tabular*}
\caption{
Differential sensitivity of the MAGIC telescopes obtained with the medium zenith angle ($30-45^\circ$) Crab Nebula data sample.
Columns as in Table~\ref{tab:diffsens}.
}\label{tab:diffsens2}
\end{table*}

\section*{Acknowledgements}
We would like to thank
the Instituto de Astrof\'{\i}sica de Canarias
for the excellent working conditions
at the Observatorio del Roque de los Muchachos in La Palma.
The support of the German BMBF and MPG,
the Italian INFN, 
the Swiss National Fund SNF,
and the Spanish MINECO %% MICINN ==> MINECO
is gratefully acknowledged.
This work was also supported
by the CPAN CSD2007-00042 and MultiDark CSD2009-00064 projects of the Spanish Consolider-Ingenio 2010 programme,
by grant 127740 of the Academy of Finland,
by the DFG Cluster of Excellence ``Origin and Structure of the Universe'',
by the Croatian Science Foundation (HrZZ) Project 09/176,
by the DFG Collaborative Research Centers SFB823/C4 and SFB876/C3,
and by the Polish MNiSzW grant 745/N-HESS-MAGIC/2010/0.
J. S. is supported by ERDF and the Spanish MINECO through FPA2012-39502 and JCI-2011-10019 grants.
We thank the two anonymous referees for their comments which helped to improve the paper.

\end{document}